\documentclass[aps,physrev,reprint,superscriptaddress]{revtex4-2}

\usepackage{import}
\usepackage{bbold}
\usepackage{amsmath}
\usepackage{tabularx}
\usepackage{graphicx}
\usepackage{hyperref}
\usepackage{color}
\usepackage{comment}
\usepackage{tikz}
\usepackage{pgfplots}
\usepackage{physics}
\usepackage{mathtools}
\usepackage{makecell}
\usepackage{multirow}

\newcommand{\av}[1]{\left\langle {#1} \right\rangle}
\newcommand\bea{\begin{eqnarray}}
\newcommand\eea{\end{eqnarray}}
\newcommand\be{\begin{equation}}
\newcommand\ee{\end{equation}}
\newcommand{\tight}[1]{\!#1\!}

\begin{document}

\title{Criticality and universality in network dismantling}

\author{Lorenzo Cirigliano}
\affiliation{\href{https://ror.org/009gyvm78}{The Abdus Salam International Centre for Theoretical Physics (ICTP)}, I-34151 Trieste, Italy}
\affiliation{Dipartimento di Fisica, Universit\`a \href{https://ror.org/02be6w209}{La Sapienza}, P.le  A. Moro, 2, I-00185 Rome, Italy}
\author{Claudio Castellano}
\affiliation{\href{https://ror.org/05rcgef49}{Istituto dei Sistemi Complessi (ISC-CNR)}, Via dei Taurini
  19, I-00185 Rome, Italy}
\author{Minsuk Kim}
\affiliation{Senseable City Laboratory, Massachusetts Institute of Technology, Cambridge, Massachusetts 02139, USA}
\affiliation{Center for Complex Networks and Systems Research, Luddy School of Informatics, Computing, and Engineering, Indiana University, Bloomington, Indiana 47408, USA}
\author{Filippo Radicchi}
\email{f.radicchi@gmail.com}
\affiliation{Center for Complex Networks and Systems Research, Luddy School of Informatics, Computing, and Engineering, Indiana University, Bloomington, Indiana 47408, USA}
\author{Hanlin Sun}
\email{hanlinsun.work@gmail.com}
\affiliation{Nordita, KTH Royal Institute of Technology and Stockholm University, Hannes Alfv\'ens v\"ag 12, SE-106 91 Stockholm, Sweden}
\affiliation{Departamento de Electromagnetismo y Física de la Materia and Instituto Carlos I de Física Teórica y Computacional, Universidad de Granada, 18071 Granada, Spain}

\begin{abstract}
Identifying the smallest set of elements whose removal dismantle a complex network, 
known as the network dismantling problem,
is a fundamental task with many practical applications. Whereas network dismantling 
has been extensively studied over the past decade, most work has focused on developing efficient algorithms for large but finite networks. By contrast, the physics of the 
network dismantling process, namely how the network structural connectivity is affected by the removal of nodes or edges,
remains largely unexplored in the thermodynamic limit. Here, we shed  light on this understudied aspect of network dismantling
by introducing an adaptive biased percolation process able to optimally dismantle a network.
% reduce network robustness.
Through a systematic analysis of synthetic network models, we find that the proposed percolation process 
displays a universal phase transition, characterized by the abrupt and simultaneous disappearance of both the giant connected component and the largest 2-core, across networks with markedly different degree distributions. Simulations on real networks further support this universality, indicating that the physics of network dismantling is insensitive to a broad range of topological properties. Together, these results suggest that a topology-agnostic theory could be developed to explain the critical behavior of network dismantling.
\end{abstract}
\maketitle

\section{Introduction}

Percolation theory studies how the existence of the macroscopic connected component in a network, often referred to as the network's giant component (GC), can be compromised by the deletion of its nodes (site percolation) or edges (bond percolation)~\cite{stauffer2018introduction}. 
The specific rules used to determine how such
microscopic elements are deleted 
define the percolation model at hand. In the ordinary percolation model, deleted microscopic elements are chosen uniformly at random.
In biased percolation models, elements' removal is decided on the basis on the network's topological properties.
Targeted percolation, consisting in the sequential removal of nodes in decreasing order of degree centrality,
is among the most celebrated examples of biased percolation models~\cite{cohen2001breakdown}, but a great variety of other biases have been investigated~\cite{holme2002attack,hooyberghs2010biased,iyer2013attack,almeira2020scaling,kim2022kselective}.
Biased percolation models can be of static or adaptive type.
In static models, biases are estimated for the
undamaged network and never updated; in adaptive models instead, biases are recalculated as the percolation process unfolds, 
often after each element's removal, so that the deletion of a specific node/edge
directly affects the subsequent removal decisions.
Another crucial distinction is due to the local or nonlocal nature of the bias, 
whether the bias is determined based on the topology of 
 a short-range neighborhood (e.g., the degree centrality) or the whole network (e.g., the betweenness centrality).

Biased percolation models are heavily used in the study of network dismantling, or so-called optimal percolation problem,
a network optimization problem that has attracted considerable interest in the past decade~\cite{morone2015influence, wandelt2018comparative, artime2024robustness}. 
Originally introduced by Morone and Makse~\cite{morone2015influence}, the problem consists in finding the smallest set of microscopic elements whose removal destroys the GC of a network.
A generalized version of the optimization problem where elements may have different cost of removal has been later introduced by Ren {\it et al.}~\cite{ren2019generalized}. The problem has been further extended to multilayer networks~\cite{osat2017optimal, baxter2018targeted, gu2025deep}. Also, two related, but different optimization problems are generally termed as network dismantling. These differ in the objective function being optimized. The most studied version of the problem consists in finding the minimum-cost set of microscopic elements whose removal is able to reduce the size of the GC of the network below a given threshold value, typically set equal to the square root of the network size~\cite{clusella2016immunization}. In the other variant of the problem, the objective function to be minimized is the robustness metric by Schneider {\it et al.}, which quantifies how quickly the size of the GC decreases as the cost of the microscopic elements that are sequentially removed from the network increases~\cite{schneider2011mitigation,shen2013discovery}.

Network dismantling is known to be a NP-hard problem~\cite{braunstein2016network}, but several highly effective strategies exist to tackle either one, the other or both variants of the problem: from the belief
propagation-guided decimation~\cite{mugisha2016identifying} and the
Min-Sum algorithms~\cite{braunstein2016network} 
to more recent approaches based on graph embeddings/machine learning~\cite{fan2020finding,grassia2021machine,chen2023searching,osat2023embedding, fu2026exploring}. 
Apart for a few exceptions, all the proposed dismantling algorithms are {\it de facto} designed to be biased and adaptive percolation models, 
where links or nodes are removed sequentially, and at each step the element to be removed is chosen greedily according to a heuristic or approximate optimization criterion evaluated on the current, partially dismantled network.
Another unifying feature of all these works on network dismantling is their primary focus on the practical aspects of the problem, i.e., how to obtain effective and efficient solutions in large, but finite networks. 
However, the actual physics of these biased adaptive percolation transitions, motivated by certain optimization criteria, occurring in infinite-size networks has received very limited attention~\cite{kim2020critical}.
The goal of the present paper is to shed some light on this understudied aspect of network dismantling processes.
We specifically achieve such a goal for bond percolation with unit cost of removal.
Our contribution is threefold.

 First, we introduce a biased percolation model where edges are sequentially removed depending on the value of the non-backtracking edge centrality (NBC), i.e., a metric proposed in this paper that depends on both the left and the right principal eigenvectors of the non-backtracking (NB) graph operator~\cite{hashimoto1989zeta}. Our decision of relying on the spectrum of the NB operator 
 stems from well-established results in the literature on network percolation: the leading eigenpair of the NB matrix is known to well characterize ordinary percolation in sparse locally tree-like networks~\cite{karrer2014percolation, hamilton2014tight}; also, centrality metrics based on the spectrum of the NB operator have been used to design well-performing algorithms for network dismantling~\cite{morone2015influence, braunstein2016network, osat2023embedding}. Here, we go beyond these studies by mathematically showing that NBC plays a crucial role in network dismantling. Indeed, we demonstrate that, in any network, removing the edge with the largest NBC causes the largest reduction, among all single-edge removals, 
 of the robustness of the network. 
 In other words, the adaptive biased percolation model that sequentially removes the largest-NBC edge corresponds to a proven quasi-optimal, greedy algorithm for network dismantling. 
 We refer to this model as the max-NBC model. Our mathematical characterization generates also a physical interpretation of the NBC as a metric quantifying the degree of participation of edges in loopy structures within the 2-core (2C) of a network. In particular, as only edges belonging to the 2C of the network have non-zero NBC, the max-NBC edge-removal protocol necessarily reduces the network to a loop-less graph (i.e., a forest or a collection of trees). This finding allows us to interpret the proposed biased percolation model as an edge-based version of the CoreHD dismantling strategy, one of the simplest and most effective algorithms to approximate solutions to the optimal site-percolation problem with unit removal cost~\cite{zdeborova2016fast, schmidt2019minimal}. Further, it  motivates us to monitor the percolation transition not just through the lens of the network's GC, but also via its largest 2C.

Second and more important, we characterize the critical properties of the network dismantling transition in random graphs with degree distributions that are either Poisson or power law. We show that both the GC and the largest 2C disappear at the same critical point, when the network actually becomes a forest. We derive an analytical expression for the critical point. Also, we perform finite-size scaling (FSS) analysis to determine the critical exponents characterizing the disappearance of the GC and of the 2C in the limit of infinitely large networks. In addition to the max-NBC model, we study softer variants of the percolation model, where the probability of removing an edge is a tunable function of  its NBC. Irrespective of the specific model variant, the GC is always characterized by a discontinuous transition. The 2C instead displays a continuous phase transition, whose abruptness depends on the specific function used to bias the selection of the edges. A major finding is that the critical exponents that characterize the disappearance of the 2C under the max-NBC edge-removal protocol do not depend on the degree distribution of the graph. The high computational cost of adaptively recalculating NBC at each stage of the percolation model, however, imposes a limit to the size of the network size that we could analyze. Taking inspiration from the similarity with CoreHD, we therefore devise a surrogate version of the biased percolation model that does not require computing the principal eigenvectors of the NB matrix. Instead, each edge within the 2C is characterized by a score given by the sum of the degrees of its incident nodes. Such a model behaves qualitatively similar to the NBC-based percolation model, but at a lower computational cost which enables the analysis of much larger network sizes.

Finally, we highlight an hyper-scaling relationship between the critical exponent values that characterize the transition of the 2C and the scaling  exponent of the diameter of the largest tree in the forest emerging at criticality. Although such a relationship is rather intuitive,
it turns out to be essential in easing the numerical estimation of the critical exponents that characterize the disappearance of the 2C. As a matter of fact, it allows us to monitor the transition of the 2C without directly measuring its size, rather by simply measuring the diameter of the tree appearing at criticality.
Even more important, the relationship allows us to non-trivially extend the FSS analysis to real networks. Our main finding is that  the disappearance of the 2C under the max-NBC edge-removal protocol is 
very similar
for all networks, regardless of their particular topological properties.

\section{Biased percolation models}
\label{sec:biased_model}

In this paper, we use a unified notation that we summarize in 
Tab.~\ref{tab:notation} 
of Appendix~\ref{app:notation} 
for ease of reference. 

\subsection{Description}

We introduce here a general modeling framework valid for all percolation models that will be considered in this paper. The input of each of these models is an unweighted and undirected network or graph $G = \left( \mathcal{N}, \mathcal{E} \right)$. $\mathcal{N}$ is the set of nodes in the graph, and $\mathcal{E}$ is the set of its edges. The sizes of these two sets are respectively denoted by $N = | \mathcal{N}|$ and $E =  |\mathcal{E}|$. 
 Consider a generic edge sequence $\vec{\epsilon} = (\epsilon_1, \epsilon_2, \ldots, \epsilon_t, \ldots, \epsilon_{E-1}, \epsilon_{E})$ out of the $E!$ possible sequences that can be constructed. Edges are removed one at a time from the graph, following the order prescribed by the sequence. Denote with $t = 0, 1, \ldots, E$ the stage of the algorithm when edge $\epsilon_t$ is removed from the graph, with the convention that, at stage $t = 0$, no actual edge is removed from the graph. Further, denote with $\mathcal{E}_t = \mathcal{E} \setminus \cup_{z = 1}^{t} \{\epsilon_z\}$ the set of edges that are still present in the graph at stage $t$ of the percolation process, so that the actual remnant graph at stage $t$ is $G_t = \left( \mathcal{N}, \mathcal{E}_t \right)$. By definition, we have that $\mathcal{E}_{0} = \mathcal{E}$ and $\mathcal{E}_{E} = \emptyset$, thus $G_0 = G$ and $G_{E} = \left(\mathcal{N}, \emptyset \right)$. 

Different percolation models differ on the specific protocol used to construct the sequence 
$\vec{\epsilon}$
of the edges to be removed. Please note that a generic percolation model can have a stochastic nature, thus the very same model can generate different edge sequences. We refer to each of those sequences as an instance of the percolation model. According to this formulation for example, an instance of the ordinary bond-percolation model is given by a random permutation of the edges in the graph.

\begin{figure*}[!htb]
    \centering
    \includegraphics[width=0.95\linewidth]{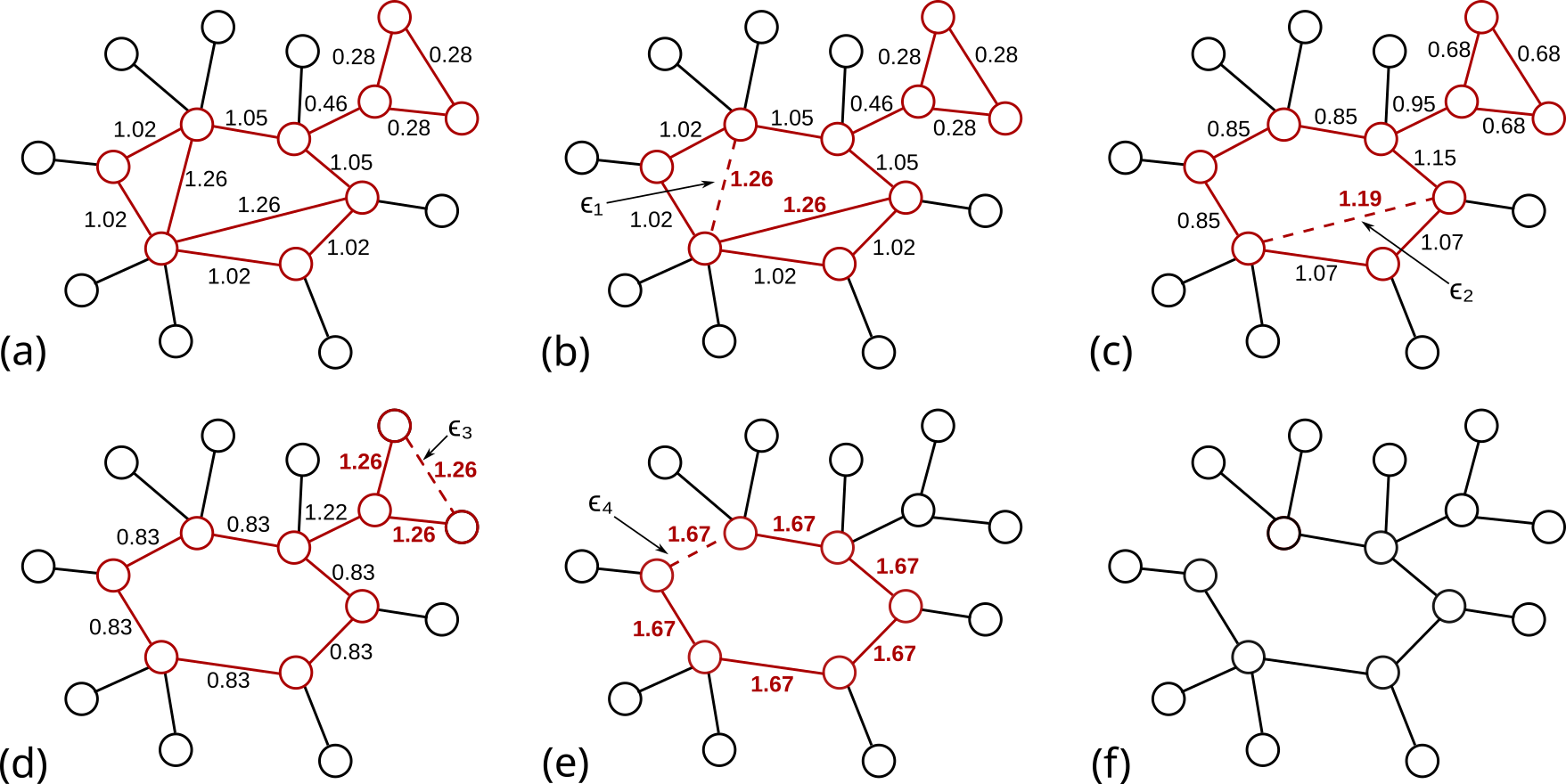}
    \caption{{\bf Biased adaptive percolation.}
    Schematic representation of a biased and adaptive percolation process, Eq.~\eqref{eq:bias}, with bias $a=+\infty$. Scores are $s(e,G_{t-1})>0$ for the edges in the $2$-core (highlighted in red), and zero otherwise; their values are reported. In this example, edge scores are equal to their non-backtracking centralities (NBC), Eq.~\eqref{eq:NBC}, (multiplied by a factor of $10$) and the process represented is the max-NBC-biased percolation described in details in Sec.~\ref{sec:NBC_biased_percolation}. (a) The graph $G_0$ with all non-zero scores reported.  Panels (b)-(e) correspond to the different stages of the process: (b) $t=1$, (c) $t=2$, (d) $t=3$, and (e) $t=4$. At each step, the edge with the highest score (bold fonts) is deleted from the graph (dashed line). Eventual ties are broken at random. The process is adaptive as the value of the scores change at each stage. (f) After the removal of $\epsilon_4$, all the scores are equal to zero and subsequent edges are removed uniformly at random.}
    \label{fig:schematic_max_NBC}
\end{figure*}

The basic mechanism used by a generic biased and adaptive percolation model to construct a sequence 
$\vec{\epsilon}$
of edges 
to be removed from the graph is as follows. At each stage $t = 1, \ldots, E$ of the construction of the sequence, each edge $e \in \mathcal{E}_{t-1}$ is associated with a score $s(e , G_{t-1}) \geq 0$, which crucially depends on the topology of the remnant graph $G_{t-1}$ to account for the adaptivity of the process.
 For non-adaptive percolation models, we have that $s(e , G_{t-1}) = s(e,G_0) = s(e)$, $\forall t = 1, \ldots, E$.  The score $s(e , G_{t-1})$ serves as a bias for the selection of the edge $e$ to become the next one in the sequence under construction. According to this formulation for example, the ordinary bond-percolation model is described by $s(e) = $ const. At stage $t$, the edge $e$ with strictly positive score $s(e, G_{t-1}) > 0$  is selected with probability
 \begin{equation}
 P\left(\epsilon_t = e 
 \right) = \frac{\left[ s(e, G_{t-1}) \right]^a} {\sum_{f \in \mathcal{E}_{t-1}} \left[ s(f, G_{t-1}) \right]^a} \; .
 \label{eq:bias}
 \end{equation}
 Here, the parameter $ - \infty \leq a \leq + \infty$ tunes the strength of the bias. For $a = 0$ for example, there is no bias and all edges are equally likely to be selected; for $a = + \infty$ ($- \infty$), the edge with maximal (minimal) score is selected deterministically; other values of the parameter $a$ bias the selection process in a probabilistic manner. However, if $s(e, G_{t-1}) = 0$ for all $e \in  \mathcal{E}_{t-1}$, then Eq.~(\ref{eq:bias}) is replaced by $ P\left(\epsilon_t = e \right) = 1/|\mathcal{E}_{t-1}|$.
 See Fig.~\ref{fig:schematic_max_NBC} for a schematic representation of a biased, adaptive percolation process. 

Given an edge sequence 
$\vec{\epsilon}$, or equivalently a series of graphs $G_0, G_1, \ldots, G_t, \ldots, G_E$,
we monitor the percolation process using two main order parameters: (i)
the fraction $S(G_t)$
of nodes belonging to the GC of the network and (ii) the fraction $C (G_t)$ of nodes belonging to the largest 2C of the network, both measured on the remnant graph  $G_t$.
Clearly, for $0 \leq t \leq E $ we have that $S(G_0) \geq  S(G_t) \geq S(G_{E}) = 1/N$ and $C(G_0) \geq  C(G_t) \geq C(G_{E}) = 0$, thus both order parameters are non-increasing functions of $t$. 

\subsection{Numerical simulations}

While studying numerically a biased percolation process $\mathrm{x}$ on a graph $G$, we denote with $_qG_t^{(\mathrm{x})}$ the graph obtained at stage $t$ in instance $q$ of process $\mathrm{x}$. The average values of the order parameters over $Q$ independent realizations are defined as 
\begin{equation}
\langle S^{(\mathrm{x})} (t, G) \rangle = \frac{1}{Q} \, \sum_{q=1}^Q \,  S(_qG_t^{(\mathrm{x})}) \; ,
 \label{eq:order_parameter}
\end{equation}
and similarly for $\langle C^{(\mathrm{x})}(t, G) \rangle$.
In the $q$-th instance, we estimate the 
pseudo-critical point as
\begin{equation}
    _{q}\hat{t}^{(\mathrm{x, GC})}(G) = \arg \max_{t} \left[ \, S(_qG_t^{(\mathrm{x})}) -  \, S(_qG_{t-1}^{(\mathrm{x})}) \right] 
    \; ,
    \label{eq:pseudo_critical}
\end{equation}
i.e., the number of edges that need to be removed in order to observe the largest drop in the size of the GC. This is a typical measurement performed while studying critical processes under the event-based ensemble~\cite{fan2020universal,li2024crossover}.
We further denote with $_q\hat{S}^\mathrm{(x, GC)}$ 
the value of the order parameter at pseudo-criticality.
The average value of the pseudo-critical point $\langle \hat{t}^{(\mathrm{x, GC})}(G) \rangle$, as well as the average value of pseudo-critical order parameter 
$\langle \hat{S}^{\mathrm{(x, GC)}}(G) \rangle$ 
is estimated using an expression similar to Eq.~(\ref{eq:order_parameter}). We also define the 
standard deviation of the pseudo-critical point as
\begin{equation}
\sigma^{(\mathrm{x, GC})}(G)  = \sqrt{\frac{1}{Q} \, \sum_{q=1}^Q \, \left[ \, _{q}\hat{t}^{(\mathrm{x, GC})}(G) \right]^2 - \left[ \langle \hat{t}^{(\mathrm{x, GC})}(G) \rangle\right]^2} \; .
    \label{eq:average_pseudo_critical}
\end{equation}
The above definitions are valid when pseudo-criticality is determined based on Eq.~(\ref{eq:pseudo_critical}). We can, however, define a similar expression for $_{q}\hat{t}^{\mathrm{(x, 2C)}}(G)$, i.e.,
\begin{equation}
    _{q}\hat{t}^{\mathrm{(x, 2C)}}(G) = \arg \max_{t} \left[ \, C(_qG_t^{(\mathrm{x})}) -  \, C(_qG_{t-1}^{(\mathrm{x})}) \right] 
    \; ,
    \label{eq:pseudo_critical_2}
\end{equation}
where the largest drop is measured in terms of 2C rather than GC. All other quantities can be defined in analogous way as well.

To make straight comparisons across networks with different number of edges, results of our numerical simulations are reported as functions of the control parameter 
\begin{equation}
    p =  \frac{t}{E}
    \; ,
\label{eq:control}
\end{equation}
which represents the fraction of edges removed from the graph. 
The change of variable is trivial, e.g.,
$\langle S^{(\mathrm{x})}(p, G) \rangle =  \langle S^{(\mathrm{x})}(t = p \, E, G) \rangle$ and $_q\hat{p}^{\mathrm{(x, GC)}}(G) = \frac{1}{E} \, _q{\hat{t}^{\mathrm{(x, GC)}}(G)}$.

\subsection{Finite-size scaling (FSS) analysis}
\label{sec:fss}

When performing finite-size scaling (FSS) analysis on network models with variable size $N$, we use a similar notation as above, but 
we replace $G$ with $N$ to highlight the dependence of the various observables from the network size $N$.  
For example, $\langle S^{(\mathrm{x})}(p, N) \rangle$ and $\langle C^{(\mathrm{x})}(p, N) \rangle$ denote the average values of the order parameters over multiple instances of the network model with size $N$ when a fraction $p$ of edges have been removed from the graphs according to protocol x. Also in FSS analyses, we tacitly assume that averages are taken over distinct instances of the network model and of the percolation process rather than simply different instances of the percolation process on the same graph. We test a few main scaling relations. 

We monitor the scaling of the pseudo-critical points using
\begin{equation}
\delta^{\mathrm{(x,o)}}(N) = \left| \langle \hat{p}^{\mathrm{(x,o)}}(N) \rangle- p_c^{\mathrm{(x,o)}}  \right|  \sim N^{-\frac{1}  {\bar{\nu}_1^{\mathrm{(x,o)}} }}
    \label{eq:scaling_pseudo1}
\end{equation}
and
\begin{equation}
\sigma^{\mathrm{(x,o)}}(N)   \sim N^{-\frac{1} { \bar{\nu}_2^{\mathrm{(x,o)}} }} \; .
    \label{eq:scaling_pseudo2}
\end{equation}
Here, o = GC, 2C to  distinguish pseudo-critical points depending on the order parameters being observed, see Eqs.~(\ref{eq:pseudo_critical}) and~(\ref{eq:pseudo_critical_2}). We stress that Eq.~(\ref{eq:scaling_pseudo1}) requires the estimate of the thermodynamic critical point $p_c^{\mathrm{(x,o)}}$ of the network model. As described later, this is obtained analytically for the network models that we consider in this paper.

We also consider the scaling of the pseudo-critical values of the order parameters

\begin{equation}
    \langle \hat{S}^{\mathrm{(x,o)}} (N) \rangle 
    \sim N^{- \frac{\beta^{\mathrm{(x,GC)}} }{ \bar{\nu}_1^{\mathrm{(x,GC)}} }}
    \label{eq:scaling_gc}
\end{equation}
and
\begin{equation}
    \langle \hat{C}^{\mathrm{(x,o)}} (N) \rangle \sim N^{-\frac{\beta^{\mathrm{(x,2C)}} }{ \bar{\nu}_1^{\mathrm{(x,2C)}} }}.
    \label{eq:scaling_2c}
\end{equation}

We finally test data collapse using the 
single-instance-based FSS ansatz~\cite{li2023explosive, li2024explosive}
\begin{equation}
    \langle S^{\mathrm{(x,GC)}}(p, N) \rangle = N^{-\frac{\beta^{\mathrm{(x,GC)}} }{ \bar{\nu}_1^{\mathrm{(x,GC)}} }} \;
    \langle V^{\mathrm{(x,GC)}} (p, N, \bar{\nu}_1^{\mathrm{(x,GC)}}) \rangle 
    \; ,
    \label{eq:collapse}
\end{equation}
where
\[
\begin{array}{l}
\langle V^{\mathrm{(x,GC)}} (p, N, \bar{\nu}_1^{\mathrm{(x,GC)}}) \rangle \\
= \frac{1}{Q} \sum_{q=1}^Q V \left[ (p - _{q}\hat{p}^{\mathrm{(x,GC)}}(N) ) \, N^{\frac{1} { \bar{\nu}_1^{\mathrm{(x,GC)}}} } \right] 
\end{array}
\; .
\]
An analogous equation is valid for $\langle C^{\mathrm{(x,2C)}}(p, N) \rangle$.
In the above equation, 
$V(\cdot)$ is a scaling function, whose details are not necessary to test data collapse.

The best estimates of the critical exponents appearing in Eqs.~(\ref{eq:scaling_pseudo1}),~(\ref{eq:scaling_pseudo2}), ~(\ref{eq:scaling_gc}) and~(\ref{eq:scaling_2c}) 
are obtained by performing simple linear regression on log-transformed variables. The fit is generally performed only on data points corresponding to network sizes $N \geq 10^3$. The goodness of the fit is quantified by the square of the Pearson correlation coefficient $0 \leq R^2 \leq 1$. The uncertainty associated to the best estimate of a critical exponent is calculated by dividing the standard deviation of the residuals by the square root of the sum of squared differences of the x-values. This is still performed on the log-transformed variables. For simplicity, we report all estimates of the critical exponents with at most two significant decimals; also, we use the value $0.01$ to report estimates of the uncertainties smaller or equal than $0.01$. 
The data collapse of Eq.~(\ref{eq:collapse}) 
is performed using the best estimates of the critical exponents
obtained via simple linear regression. The quality of a collapse is assessed only visually.

\subsection{Connection to network dismantling}

We highlight here the connection between our model's formulation and the more general problem of network dismantling. 
Following the typical convention used in network dismantling, a finite network with size $N$ is considered dismantled when the size of its giant component is smaller than $ \sqrt{N}$~\cite{clusella2016immunization}. Given an edge sequence 
$\vec{\epsilon}$,
one identifies the smallest value of $p$, namely 
$\tilde{p}(\vec{\epsilon})$,
such that 
the resulting GC is smaller than the above threshold.
The optimal bond-percolation problem with unit-cost removal then consists in finding the edge sequence 
$\vec{\epsilon}^*$
with the smallest 
$\tilde{p}(\vec{\epsilon})$
value. 
The exact solution to the problem requires testing all $E!$ permutations of the graph edges, but clearly such a brute-force approach is feasible only in ultra-small networks. In general networks, one can attempt to approximate solutions to the problem by implementing computationally feasible ways of building effective edge sequences.
In the next section, we introduce a specific biased percolation model that provides a greedy quasi-optimal solution to the dismantling problem.

\section{Non-backtracking centrality (NBC)}

%In percolation theory, the pseudo-critical threshold of the ordinary percolation model on a graph is  closely associated with the Hashimoto non-backtracking (NB) matrix of the graph ~\cite{hashimoto1989zeta}. 
Given an undirected graph $G$ with edge set $\mathcal{E}$, we define the directed edge set $\overline{\mathcal{E}}= \{ i \tight{\to}j, j \tight{\to} i:(i,j)\in \mathcal{E} \}$, with $|\overline{\mathcal{E}}|=2E$. 
%The NB matrix 
The
Hashimoto non-backtracking (NB) matrix 
$B(G)$ of the graph $G$ is a $2E \times 2E$ matrix defined on the elements of the directed edge set $\overline{\mathcal{E}}$~\cite{hashimoto1989zeta}. 
Given two directed edges $i \to j$ and  $k \to l$, the corresponding element of the NB matrix is
\begin{equation}
    B_{i \to j, k \to l} (G) = 
    \delta_{j,k}
    (1-\delta_{i,l}) 
    \, ,
    \label{eq:matrix_elements_B}
\end{equation}
where we used 
the Kronecker delta function $\delta_{x, y} = 1$ if $x = y$ and $\delta_{x,y} = 0$ otherwise.

The principal eigenvalue $\Lambda(G)$ of $B(G)$ is a fundamentally important indicator of networks' robustness~\cite{karrer2014percolation,hamilton2014tight,timar2017nonbacktracking}. For an arbitrary graph $G$, we can write
\bea
\langle p_c^{\mathrm{(ORD, GC)}}(G) \rangle \leq 1 - \frac{1}{\Lambda(G)} \; ,
\label{eq:upper_bound}
\eea 
meaning that the quantity $1-1/\Lambda(G)$ is an upper bound for the pseudo-critical point of ordinary bond percolation, and the bound is tight for large and locally tree-like graphs~\cite{radicchi2015predicting}.
The left and right eigenvectors associated to $\Lambda$, also called Perron eigenvectors, play a crucial role in our analysis. They are defined by $\bra{\ell}B=\Lambda \bra{\ell}$ and $B\ket{r}=\Lambda \ket{r}$, respectively, and if $\Lambda>1$ they can be chosen with non-negative entries and normalized such that $\braket{\ell}{r}=1$.

We are interested in how the NB leading eigenvalue of a graph
evolves as its edges are sequentially removed.
Consider the removal of the edge $e = (i, j)$ from the graph $G = (\mathcal{N}, \mathcal{E})$ that leads to the graph $G \setminus e = (\mathcal{N}, \mathcal{E} \setminus \{e\})$. As detailed in Appendix~\ref{appendix:perturbation}, this is a rank-2 perturbation on the NB matrix of the graph, which naturally decreases the value of the corresponding principal NB eigenvalue. At leading order,  the eigenvalue change $\delta \Lambda (e,G) = \Lambda(G\setminus e)-\Lambda(G)$ 
is
\begin{equation}
    \delta \Lambda(e,G) \simeq -\Lambda(G) h(e,G), 
    \label{eq:var_lambda}
\end{equation}
where 
\begin{equation}
    h(e,G) = \ell_{i \to j}r_{i \to j} + \ell_{j \to i}r_{j \to i}.
    \label{eq:NBC}
\end{equation}
for the removal of the edge $e=(i,j)$, where $r_{i \to j}$ and $\ell_{i \to j}$ denote the components associated to the directed edge $i \tight{\to} j$ of $\ket{r}$ and $\ket{\ell}$, respectively. This motivates us to define the non-backtracking centrality (NBC) of the edge $e=(i,j)$ in the graph $G$ as $h(e,G)$. Eq.~\eqref{eq:var_lambda} shows that, at leading order, removing the edges with largest NBC centrality produces the strongest reduction in the leading NB eigenvalue (see Figure \ref{fig:NBC_vs_others}). 
Note that related NB centrality measures have appeared in previous studies~\cite{zhang2015nonbacktracking,morone2016collective}; however, the corresponding estimates of the eigenvalue shift, Eq.~\eqref{eq:var_lambda}, were not correct.

As shown in Ref.~\cite{torres2020non}, either $r_{i \to j}$ or $r_{j \to i}$ vanish for an edge $e=(i,j)$ that does not belong to the 2C of $G$. It follows that $h(e, G) > 0$ only on the 2C of $G$; otherwise $h(e,G)=0$.
Moreover, the value of the NBC metric of an edge $e$ is related to the participation of $e$ in loopy structures (i.e., their number, their length) in the graph, see  Appendix~\ref{appendix:NBC_loops} for details. In particular, the higher $h(e,G)$ is, the more loops $e$ belongs to.

\section{NBC-biased percolation}
\label{sec:NBC_biased_percolation}
The NBC-biased percolation model is obtained by setting $s(e, G_{t-1}) = h(e, G_{t-1})$ in the general framework introduced in Section~\ref{sec:biased_model}. 

Eqs.~(\ref{eq:var_lambda}) and~(\ref{eq:NBC}) tell us that, if we want to decrease as much as possible the principal NB eigenvalue 
of a graph
by removing only one edge from it, then we should select the edge with the largest NBC by setting $a = + \infty$. We refer to this special case as the max-NBC-biased percolation model. See Fig.~\ref{fig:schematic_max_NBC} for an example of the max-NBC-biased removal on a small network. 
In turn, 
given its optimality in reducing the NB leading eigenvalue
and
the validity of Eq.~(\ref{eq:upper_bound}), the  max-NBC-biased protocol 
is a greedy-optimal strategy of dismantling the graph
% reducing the robustness of the graph 
via edge removal. 

\begin{figure}[!htb]
    \centering
    \includegraphics[width=0.9\linewidth]{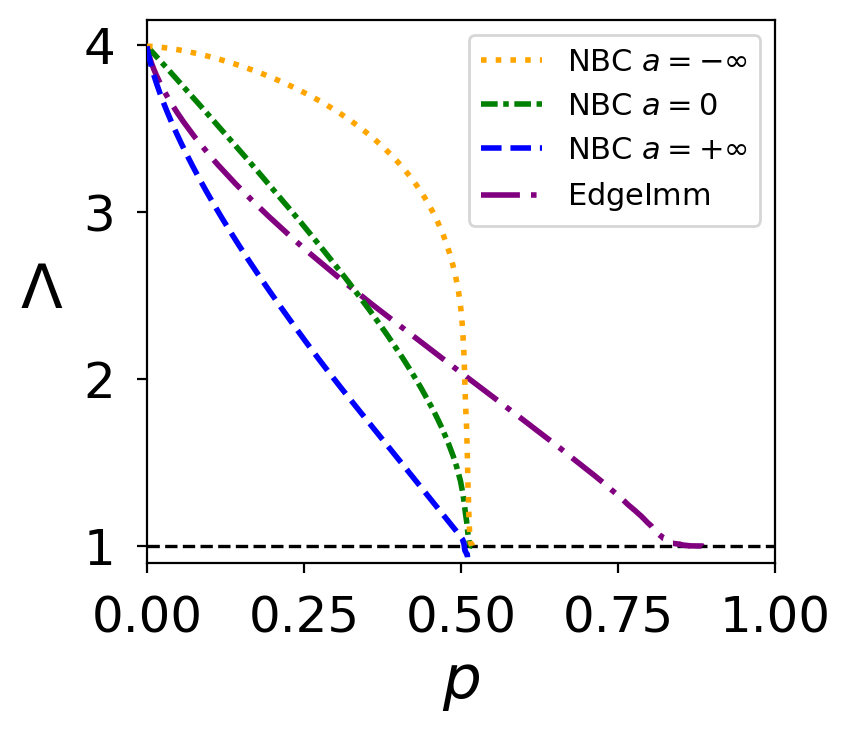}
    \caption{{\bf Quasi-optimality of NBC-biased percolation.} 
    The leading eigenvalue $\Lambda$ of the NB matrix obtained via different edge-removal strategies is shown as a function of the fraction of removed edges $p$. Link removal strategies NBC with different bias parameter $a$ are compared. We also show the benchmark strategy ``EdgeImm'' proposed in Ref.~\cite{zhang2021minimizing}.
    Each curve is obtained by averaging results over $Q=100$  instances of the Erd\H{o}s--R\'enyi model with size $N=10^4$ and  average degree $\langle k \rangle=4$.}
    \label{fig:NBC_vs_others}
\end{figure}

We now characterize the physics of such a percolation process both theoretically and numerically.

From the theoretical point of view, we can approximate
the pseudo-critical point 
$\langle \hat{p}^{\mathrm{(NBC, GC)}}(G) \rangle$ of NBC-biased percolation, using the fact that we are only removing edges that belong to 2C components.
Let $n_C$ denote the number of connected components in $G$. The cyclomatic number of $G$, also called first Betti number, is $E+n_C-N$. This quantity is the minimum number of edges to be removed to make $G$ a forest. 
The pseudo-critical point is therefore bounded from below by
\begin{equation}
    \langle \hat{p}^{\mathrm{(NBC, GC)}}(G) \rangle \geq  1 -\frac{N-n_C}{E}.
    \label{eq:feedback_edge_set}
\end{equation}
The inequality is strict for a finite graph $G$, because the NBC protocol may remove a bridge in the 2C of $G$ -- an edge with nonzero NBC that nonetheless belongs to no cycle. Such a removal does not change the cyclomatic number, so more than $E+n_C-N$ edges need to be removed before the graph becomes a forest.
Once the forest is reached, the remaining edges are removed in random order and the giant component disappears abruptly. The equality in Eq.~\eqref{eq:feedback_edge_set} holds whenever the 2C bridges are a vanishing fraction of the edges, which is the case for random networks in the infinite-size limit. 

\begin{figure*}[!htb]
    \centering
    \includegraphics[width=0.99\linewidth]{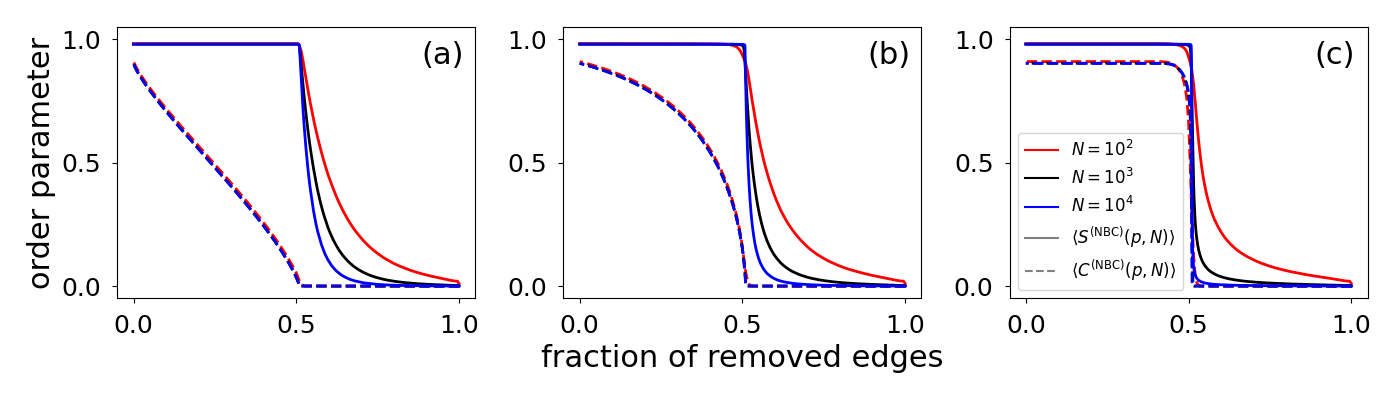}
    \caption{{\bf NBC-biased percolation on Erd\H{o}s-R\'enyi graphs.} (a) Relative size of the giant component $\langle S^{\mathrm{(NBC)}} (p, N) \rangle$
    and of the 2-core 
    $\langle C^{\mathrm{(NBC)}} (p, N) \rangle$ as a function of the fraction $p$ of edges removed from the initial graph. The initial graphs are instances of the Erd\H{o}s-R\'enyi (ER) model with average degree $\av{k} =4$ and different number of nodes $N$. Here, edges are selected based on the NBC-biased protocol with bias parameter $a = - \infty$. Results are averaged over $Q = 10000$ graph instances. (b) Same  as in (a), but with $a = 0$. (c) Same as in (a) and (b), but for $a = + \infty$.}
    \label{fig:1}
\end{figure*}

We numerically validate in Fig.~\ref{fig:1}
our claim regarding the independence of the value of the pseudo-critical point  from the bias parameter $a$. 
Here, we apply the NBC-biased percolation model to instances of Erd\H{o}s--R\'enyi (ER) graphs with average degree $\langle k \rangle = 4$ and variable size $N$.  Fig.~\ref{fig:1} shows the two order parameters $\langle S^{\mathrm{(NBC)}} (p, N) \rangle$ and
$\langle C^{\mathrm{(NBC)}} (p, N) \rangle$ as functions of $p$. We consider different values of $N$ and well as different values of the bias parameter $a$. For all values of the bias parameter $a$, the GC of the graph is left untouched by the
NBC-biased process until criticality; then, it undergoes an abrupt transition. The 2C instead shows a less trivial behavior, undergoing a smooth transition, which becomes progressively sharper as $a$ increases. We note that the rhs of Eq.~(\ref{eq:feedback_edge_set}) is a good lower bound also for $ \langle \hat{p}^{\mathrm{(NBC, 2C)}}(G) \rangle$.

\begin{figure*}[!htb]
    \centering
    \includegraphics[width=0.99\linewidth]{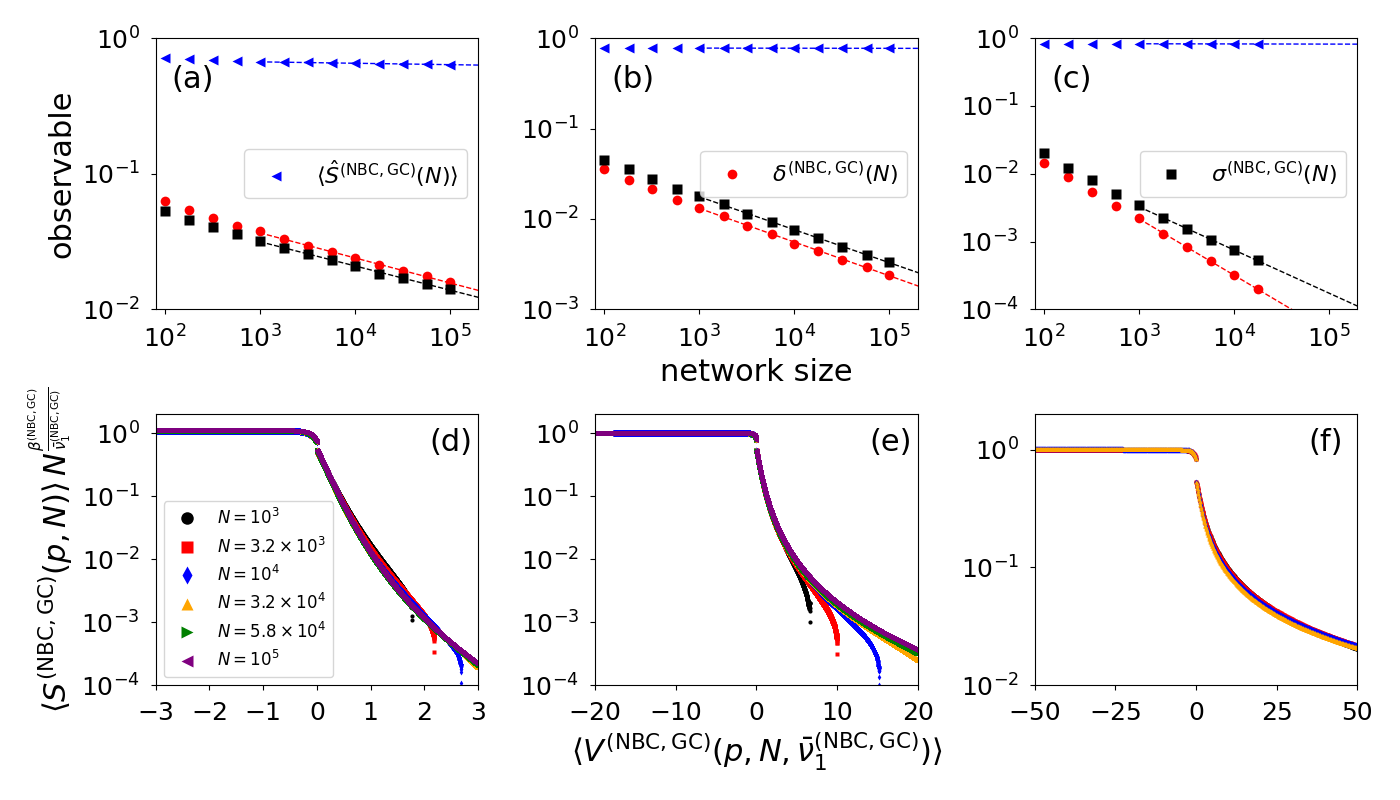}
    \caption{{\bf Critical properties of the GC transition for NBC-biased percolation on Erd\H{o}s-R\'enyi graphs.} (a) We test the scaligns of Eq.~(\ref{eq:scaling_pseudo1}), ~(\ref{eq:scaling_pseudo2}) and~(\ref{eq:scaling_gc}) for NBC-biased percolation with $a = - \infty$.
    Averages are taken over $Q = 10000$ instances of the biased percolation model. 
    The slopes of the dashed lines identify the scaling exponents. Best estimates are reported in Tab.~\ref{tab:exponents}.
    (b) and (c) Same as in panel (a), but for $a = 0$ and $a = + \infty$, respectively. 
    (d-e) We test the finite-size scaling relationship of Eq.~(\ref{eq:collapse}) for the NBC-biased percolation with bias parameter respectively equal to $a= - \infty, 0$ and $+\infty$.
    }
    \label{fig:2}
\end{figure*}

Next, we perform in Fig.~\ref{fig:2} a systematic numerical investigation of the critical properties of 
the NBC-biased percolation process on ER graphs via FSS analysis, see Sec.~\ref{sec:fss} for details.
As derived in Appendix~\ref{appendix:threshold}, 
the lower bound appearing on the rhs of Eq.~(\ref{eq:feedback_edge_set}) can be estimated for infinitely large random network as
\begin{equation}
    p_c^{\mathrm{(NBC, o)}} = 1-\frac{2-2g_0(z)+\langle k \rangle z^2}{\langle k \rangle} \; ,
    \label{eq:threshold}
\end{equation}
for both o = GC and 2C. Here, $g_0(z) = \sum_k P(k) z^k$ is the generating function of the degree distribution $P(k)$, and $z$ denotes the probability that a randomly chosen link does not lead to the GC, which satisfies the recursive equation 
\begin{equation}
z=\frac{g_0'(z)}{\langle k \rangle} .   
\label{eq:recursive}
\end{equation}
We use the expression of Eq.~(\ref{eq:threshold}) as the estimate of the thermodynamic critical point needed to test the scaling of Eq.~(\ref{eq:scaling_pseudo1}). For the specific case of ER graphs with average degree $\langle k \rangle =4$, we have that $p_c^{\mathrm{(NBC,o)}} \simeq 0.5095205748$. 
Numerical estimates of the critical exponents are reported in Tab.~\ref{tab:exponents}; we stress that those estimates are qualitative, as the network sizes that we could analyze are quite small.   As the results of Fig.~\ref{fig:2}(a-c) indicate, we find that $\bar{\nu}^{\mathrm{(NBC, GC)}}_1 = \bar{\nu}^{\mathrm{(NBC, GC)}}_2$ for $ a= -\infty, 0$, but  $\bar{\nu}^{\mathrm{(NBC, GC)}}_1 \neq \bar{\nu}^{\mathrm{(NBC, GC)}}_2$ for $ a= +\infty$.  By assuming $\beta^{\mathrm{(NBC, GC)}} = 0$, those exponent values lead also to decent data collapses, see Fig.~\ref{fig:2}(d-f). 

\begin{figure*}[!htb]
    \centering
    \includegraphics[width=0.99\linewidth]{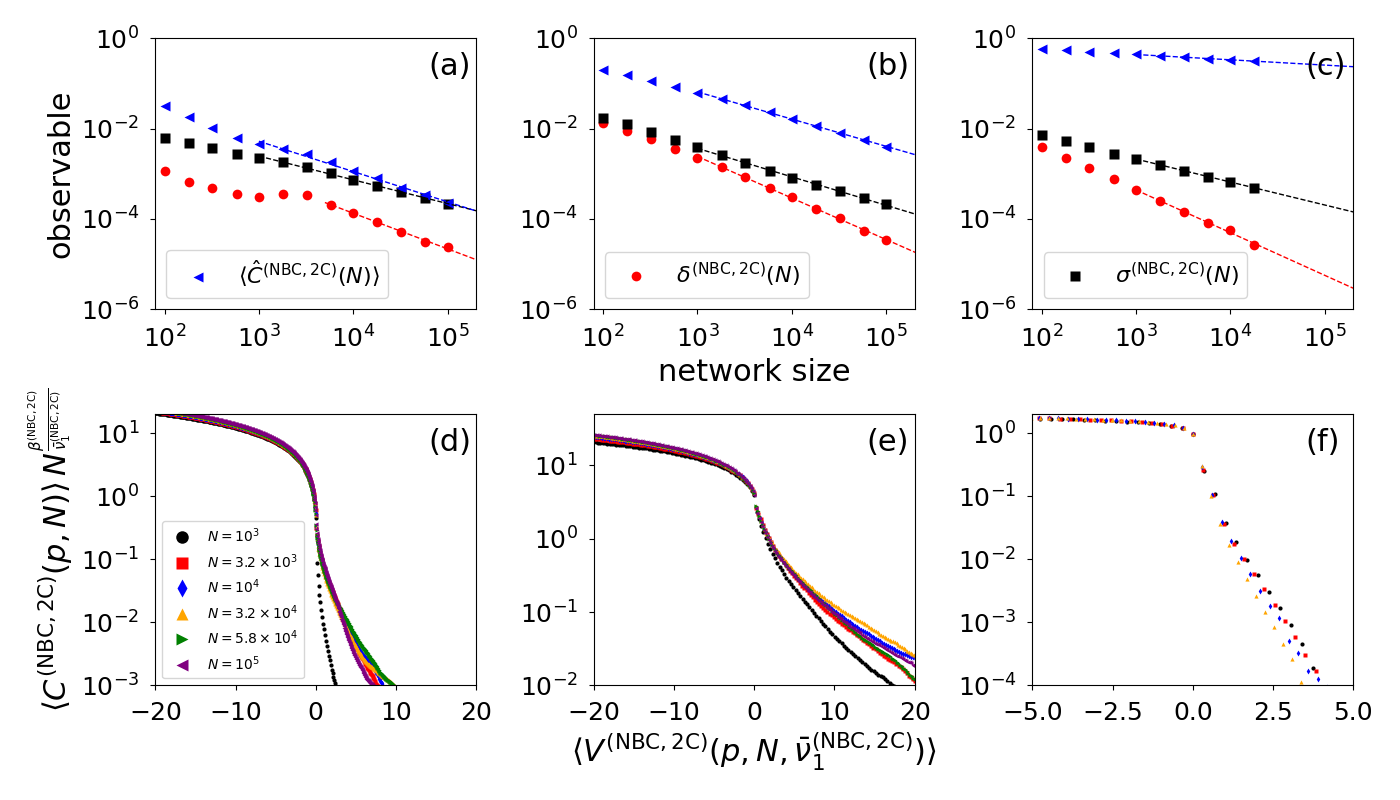}
    \caption{{\bf Critical properties of the 2C transition for NBC-biased percolation on Erd\H{o}s-R\'enyi graphs.} Same as in Fig.~\ref{fig:2}, but for the 2C transition. The best 
    of the critical exponents
    are reported in Tab.~\ref{tab:exponents}.
    }
    \label{fig:3}
\end{figure*}

We repeat in Fig.~\ref{fig:3} the FSS analysis by focusing on the size of the 2C instead of the GC. Numerical estimates of the exponents are reported in Tab.~\ref{tab:exponents}.
In this case, ${\bar \nu}^{\mathrm{(NBC, 2C)}}_1 \neq {\bar \nu}^{\mathrm{(NBC, 2C)}}_2$ for all values of $a$.
The exponent $\beta^{\mathrm{(NBC, 2C)}}$ 
is nontrivial and is a decreasing function of $a$.
For $a=0$, i.e., when edges belonging to the 2C are removed at random, the ratio
$\beta/{\bar \nu}_1^{\mathrm{(NBC, 2C)}} \simeq 2/3$.
This is the same ratio found for standard 2C percolation~\cite{dorogovtsev2020kcore}, but remarkably
NBC-biased percolation belongs to a different universality class, as for standard 2C $\beta^{\mathrm{(2C, 2C)}}=2$ and ${\bar \nu}_1^{\mathrm{(2C, 2C)}} =3$,
while here $\beta^{\mathrm{(NBC, 2C)}} \simeq 2/3$ and ${\bar \nu}_1^{\mathrm{(NBC, 2C)}} \simeq 1$.

We also studied NBC-biased percolation on random graphs with power-law degree distribution 
$P(k) \sim k^{-\gamma}$, see Figs.~\ref{fig:4old} and~\ref{fig:old5} in Appendix~\ref{app:SF}.
Similarly to what observed in other percolation processes~\cite{cirigliano2024scaling}, also the critical properties of NBC-biased percolation are expected to be 
dependent on the degree distribution of the network model.
We indeed observe that the critical exponents generally depend on the degree exponent $\gamma$.
There are only two major exceptions. First, we always find that $\beta^{\mathrm{(NBC, GC)}} = 0.00$ for the GC transition. 
Second and more surprising, we find that in the 2C transition $\beta^{\mathrm{(NBC, 2C)}} \simeq 0.10$ for any $\gamma$ as long as $a = + \infty$, see Figs.~\ref{fig:4} and ~\ref{fig:5}. 
This result matches with what observed in ER graphs, indicating that removing edges according to the max-NBC protocol homogenizes the network's topology before the transition takes place. 

Figs.~\ref{fig:2},~\ref{fig:3} and~\ref{fig:4},~\ref{fig:5} suggest that NBC-biased percolation has critical properties qualitatively different from standard 2C percolation, with exponents depending continuously on $a$, 
but never coinciding with the standard ones. The strength of this conclusion is limited by the relatively small
range of system sizes considered. This is a consequence of the adaptive and nonlocal nature of the process considered.
After each removal it is necessary to recalculate the NBC score of all edges: 
the computational complexity scales with $N$ to a power larger than 2, making the analysis of very large systems prohibitive.

\begin{table}[!htb]
\begin{tabular}{lrrrrrr}
network & a & o & $\beta$ & $\bar{\nu}_1$ &  $\bar{\nu}_2$ \\ \hline
ER & $-\infty$ & GC &  $0.06 \pm 0.01$ & $5.4 \pm 0.1$ & $5.6 \pm 0.1$\\
ER & $0$ & GC &  $0.00 \pm 0.01$ & $2.69 \pm 0.04$ & $2.75 \pm 0.03$ \\
ER & $+\infty$ & GC & $0.00 \pm 0.01$ & $1.19 \pm 0.01$ & $1.65 \pm 0.03$\\
\hline
SF $\gamma=2.50$ & $+\infty$ & GC & 0.00 $\pm$ 0.01 & 1.09 $\pm$ 0.01 & 1.06 $\pm$ 0.01  \\ 
SF $\gamma=3.50$ & $+\infty$ & GC & 0.00 $\pm$ 0.01 & 1.17 $\pm$ 0.01 & 1.13 $\pm$ 0.01  \\ 
\hline
ER & $-\infty$ & 2C & $0.85 \pm 0.06$ & $1.27 \pm 0.05$ & $1.90 \pm 0.04$ \\
ER & $0$       & 2C & $0.66 \pm 0.03$ & $1.09 \pm 0.01$ & $1.59 \pm 0.03$ \\
ER & $+\infty$ & 2C & $0.12 \pm 0.01$ & $0.99 \pm 0.04$  & $1.98 \pm 0.01$ \\
\hline
SF $\gamma=2.50$ & $+\infty$ & 2C & 0.11 $\pm$ 0.01 & 0.87 $\pm$ 0.03 & 0.92 $\pm$ 0.01  \\ 
SF $\gamma=3.50$ & $+\infty$ & 2C & 0.12 $\pm$ 0.01 & 0.96 $\pm$ 0.03 & 0.98 $\pm$ 0.01 \\ 
\hline
\end{tabular}
\caption{{\bf Critical exponents of the network dismantling transition.} We list here the best estimates of the critical exponents characterizing the NBC-biased percolation processes considered in this paper.
Results are obtained via finite-size scaling analysis on Erd\H{o}s-R\'enyi (ER) and Scale-free (SF) graphs. From left to right, we list the network model, the value of the bias parameter $a$, the specific order parameter monitored, and the corresponding values of the critical exponents characterizing the percolation transition. 
}
\label{tab:exponents}
\end{table}

Before closing this section, we would like to comment on the quasi-optimality of the proposed max-NBC strategy. While removing the largest-NBC edge is indeed optimal at leading order, there is a wide class of link removal strategies that solve the network dismantling problem equally well, producing the same percolation threshold of the rhs of Eq.~\eqref{eq:feedback_edge_set} as the max-NBC strategy does. This is due to the fact that the optimal way of damaging the GC via link removal is to damage all cycles in the network and reduce the network to a forest~\cite{karp1972reducibility, dinneen2001forbidden} and the fraction of edges that need to be removed is given by the rhs of Eq.~\eqref{eq:feedback_edge_set}. In other words, all adaptive de-cycling algorithms produce the same critical threshold in the infinite-network limit. However, the degeneracy concerns the location of the critical transition alone. The trajectory of how the network is dismantled is not degenerate, and different
de-cycling protocols yield markedly different $\Lambda$ {\it vs.} $p$ curves, see Fig.~\ref{fig:NBC_vs_others}, as well as different critical exponents for the GC and 2C transitions. We here focus on adaptive NBC-biased percolation for two reasons. First, adaptive max-NBC algorithm provides a first order approximation of a well-defined global objective of the network dismantling problem. Second, it embodies the design principle underlying many dismantling heuristics, 
namely the greedy optimization of a certain objective at each single-edge removal.

\section{2-core-degree-biased (2CDC-biased) percolation}

Is it possible to find a process with the phenomenology analogous to NBC-biased percolation but more
easily amenable to large-scale networks?
A local quantity correlated with the NBC of 
an edge is the sum of the degrees of the nodes adjacent to the edge~\cite{pastor2020localization}.
This prompts us to define the score of the edge $e = (i,j)$ as
\begin{equation}
 d ( e = (i,j), G )  = \, ^{\mathrm{(2C)}}k_i(G) + \, ^{\mathrm{(2C)}}k_j(G)  \; ,
    \label{eq:degree}
\end{equation}
where $^{\mathrm{(2C)}}k_i(G)$ and $^{\mathrm{(2C)}}k_j(G)$ are the degree of nodes $i$ and $j$ of the 2C of the graph $G$. We name the score of Eq.~(\ref{eq:degree}) as the 2-core-degree centrality (2CDC) of edge $e$. The 2CDC-biased  percolation model is then defined by setting $s(e, G_{t-1}) = d(e, G_{t-1})$ in the general framework introduced in Section~\ref{sec:biased_model}.
Note that, although the score of Eq.~(\ref{eq:degree}) is a local quantity, the calculation is performed within the 2C subgraph, whose determination requires the consideration of the whole network topology. For this
reason 2CDC-biased percolation is a genuinely nonlocal model.
 At the same time, we also remark that the determination of the 2C of a graph requires nonlocal computations only 
 a few times during the percolation process. One of such a nonlocal computation is required at the beginning.
 After that, since only one edge is removed at a time and the remaining edges can only exit the 2C, one can keep track of the 2C of the graph in a quite efficient manner by performing local operations only, or just a few global updates. 
 In short, the 2CDC-biased percolation model represents a good surrogate of the NBC-biased one, with the  computational advantage of being applicable to much larger network sizes.
 In particular, when the bias parameter is $a=0$, the two models are exactly the same. For the extreme case $a= +\infty$, we expect the 2CDC-biased percolation model to be less effective than the NBC-biased percolation model in performing network dismantling, see Fig.~\ref{fig:NBC_SM}.
Note also that $\langle \hat{p}^{\mathrm{(2CDC, x)}}(G) \rangle \simeq \langle \hat{p}^{\mathrm{(NBC, x)}}(G) \rangle$, thus we can re-use Eqs.~(\ref{eq:feedback_edge_set}) and~(\ref{eq:threshold}) to estimate the pseudo-critical and the critical point of 2CDC-biased percolation.

The results reported in Figures~\ref{fig:degree_2CDCGC} and
\ref{fig:degree_2CDCcore} fully confirm our intuition. Critical exponent values are summarized in Tab.~\ref{tab:exponents}. For $a=0$, results from the 2CDC- and NBC-biased percolation models are the same within error bars. For the GC transition, similarly to what observed in NBC-biased percolation, also for 2CDC-biased percolation we find: (i) for any $a$ value, the transition is discontinuous, characterized by a null critical exponent $\beta^{\mathrm{(2CDC, GC)}}$; 
(ii) the other critical exponents are such that $\bar{\nu}^{\mathrm{(2CDC, GC)}}_1 \simeq \bar{\nu}^{\mathrm{(2CDC, GC)}}_2$ for $a = -\infty$, but $\bar{\nu}^{\mathrm{(2CDC, GC)}}_1 \neq \bar{\nu}^{\mathrm{(2CDC, GC)}}_2$ for $a = +\infty$. For the 2C transition, all critical exponents are bias-dependent as they are for NBC-biased percolation. In particular, we always observe $\bar{\nu}^{\mathrm{(2CDC, 2C)}}_1 \neq \bar{\nu}^{\mathrm{(2CDC, 2C)}}_2$ for any $a$ value.

\section{Critical tree structure}
\label{sec:tree}

As discussed in Sec.~\ref{sec:NBC_biased_percolation}, at criticality the network reduces to a forest. The structure of this critical forest
provides valuable insights into the nature of the NBC- and 2CDC-biased
percolation processes. Since both processes remove only links from loopy structures within the graph and almost no node is disconnected from the rest of the network, the largest tree within the critical forest has essentially the same size as the original network. The specific structure of  such a tree, however, may differ substantially between the various
biased percolation processes.

\subsection{Synthetic networks}

\begin{figure*}[!htb]
    \centering
    \includegraphics[width=0.99\linewidth]{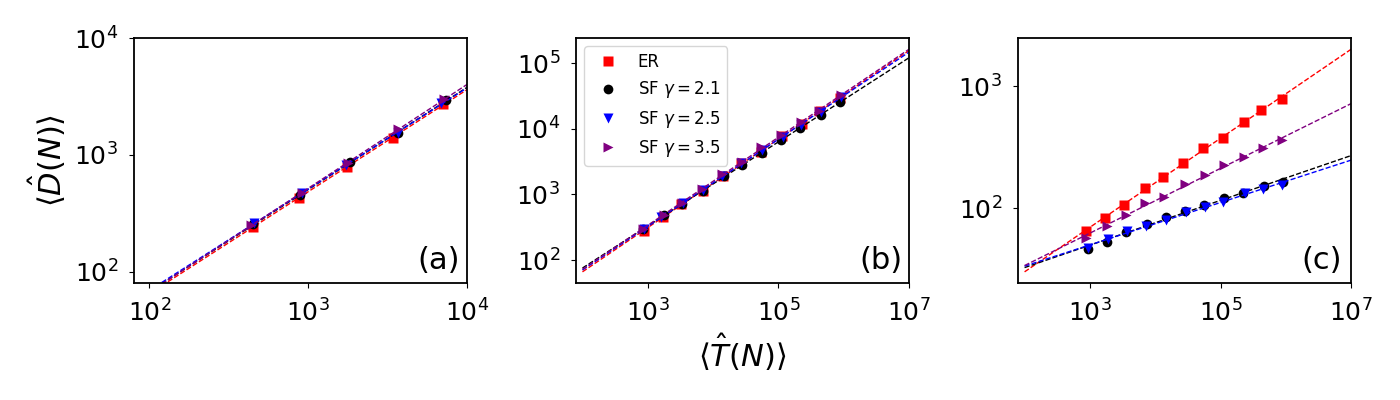}
    \caption{
    {\bf Critical properties of the NBC- and 2CDC-biased percolation transitions on synthetic graphs.}
    (a) Average value of the diameter $\langle \hat{D}(N) \rangle$ as a function of the average value of the size $\langle \hat{T}(N)\rangle$ of the largest trees in the critical forests of the NBC-biased percolation process. Here, the biased parameter is $a = +\infty$, and the percolation process is applied to instances of the  
    Erd\H{o}s-R\'enyi (ER) with size $N$ and average degree $\langle k \rangle = 4$, or scale free (SF) graphs with size $N$,  degree exponent $\gamma$, minimum degree $k_{\min}=3$, and maximum degree $k_{\max}=\sqrt{N}$. Results are averaged over at least $Q=50$ instances of the network models for given $N$. The dashed lines are best fits of the scaling Eq.~(\ref{eq:alpha}). Best estimates of the exponents are reported in Tab.~\ref{tab:exponents_tree}. (b)
    Same as in (a), but for the 2CDC-biased percolation model with $a = +\infty$.
    (c) Same as in (a), but for $a = 0$.
    }
    \label{fig:4}
\end{figure*}

We perform systematic numerical tests of our biased percolation models on both ER and SF random graphs. For instance $q$ of process x = NBC, 2CDC applied to a graph with $N$ nodes,  we characterize the pseudo-critical tree by its diameter $_q\hat{D}^{(\mathrm{x})} (N)$ and its size $_q\hat{T}^{(\mathrm{x})}(N)$. As Fig.~\ref{fig:4} shows, we find that the value of such a diameter, averaged over independent realizations of the process x, scales with the average number of nodes within the critical tree as
\bea
\langle \hat{D}^{(\mathrm{x})} (N) \rangle \sim  \langle \hat{T}^{(\mathrm{x})} (N) \rangle^{\alpha^{(\mathrm{x})}} \; .
\label{eq:alpha}
\eea

Interestingly, both max-NBC and max-2CDC percolation display universal scaling across synthetic networks with very different degree distributions. 
We estimate compatible exponent values for rather different network models, see Tab.~\ref{tab:exponents_tree}. Within error bars, these corresponds to
$\alpha^{\mathrm{(NBC)}} \simeq 0.88$ for max-NBC-biased percolation and
$\alpha^{\mathrm{(2CDC)}} \simeq 0.66$ for max-2CDC-biased percolation.
 This universality is lost when the bias parameter $a$ is small, for example for $ a= 0$, see Fig.~\ref{fig:4}(c) and Tab.~\ref{tab:exponents_tree}. In this case, details of the network model matter: the scaling of Eq.~(\ref{eq:alpha}) still holds for a network model with a given set of parameter values, 
however, each of these configurations has associated a different value of the critical exponent.

\begin{figure*}[!htb]
    \centering
    \includegraphics[width=0.99\linewidth]{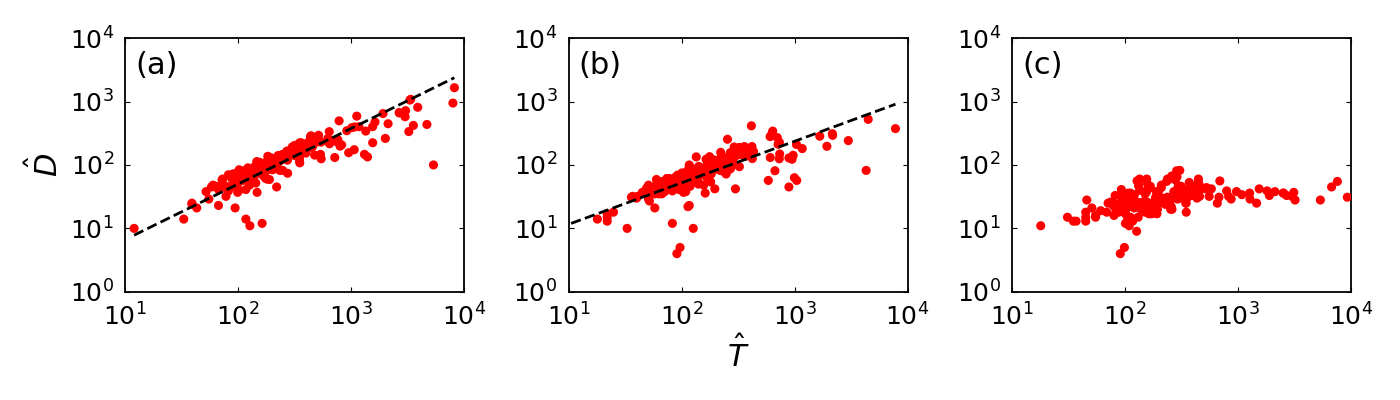}
    \caption{
    {\bf Critical properties of the NBC- and 2CDC-biased percolation transitions on real graphs.}
    Similar to Figure \ref{fig:4}(a-c) but the calculations are conducted on $217$ real networks extracted from the Netzschleuder dataset \cite{peixoto2020netzschleuder} as of August 2026. We restrict ourselves to monopartite networks with $N<10^4$ nodes and $E<10^5$ links. Directed networks are symmetrized to undirected networks, self-loops and multilinks are removed. The dashed lines in (a) and (b) corresponds to the scaling $\hat{D} \sim \hat{T}^{\alpha}$. We use $\alpha = 0.88$ in (a) and $\alpha = 0.65$ in (b), i.e., the average values of the exponents measured in Figures \ref{fig:4}(a) and (b), respectively.
    }
    \label{fig:5}
\end{figure*}

Furthermore, the critical exponent $\alpha^{(\mathrm{x})}$ is related to the critical
exponents $\beta^{\mathrm{(x, 2C)}}$ and $\bar{\nu}_1^{\mathrm{(x, 2C)}}$ of the 2C transition through the
hyperscaling relation
\bea
\alpha^{(\mathrm{x})} + \frac{\beta^{\mathrm{(x, 2C)}}}{\bar{\nu}_1^{\mathrm{(x, 2C)}}} = 1,
\label{eq:hyperscaling}
\eea
which holds for both NBC- and 2CDC-biased percolation for any value of the bias parameter $a$. Equation~(\ref{eq:hyperscaling}) admits a simple geometric interpretation. The pseudo-critical network is effectively composed of a quasi-one-dimensional 2C, i.e., its diameter grows linearly with its size $\langle \hat{D}^{(\mathrm{x})} \rangle \sim N\, \langle \hat{C}^{\mathrm{(x, 2C)}} \rangle $, and dangling trees around such a core. The two components play sharply complementary roles: the dangling trees carry almost the entire mass of the network, as the pseudo-critical 2C contains only a vanishing fraction of the network nodes, i.e., $\beta^{\mathrm{(x, 2C)}} > 0$ in Eq.~(\ref{eq:scaling_2c}), while the pseudo-critical GC contains a non-vanishing fraction of the network nodes, i.e., $\beta^{\mathrm{(x, GC)}} = 0$ in Eq.~(\ref{eq:scaling_gc}).
Eq.~(\ref{eq:hyperscaling}) provides a way of dramatically easing the analysis of the 2C transition, as it allows us to determine some of its critical properties by simply measuring the diameter of a tree instead of monitoring the evolution of the 2C of the network as the biased percolation process unfolds.

\begin{table}[!htb]
\begin{tabular}{lrrrr}
network model & x & a & $\alpha$ &  $R^2$\\ \hline
ER & NBC & $+\infty$ &  $0.87 \pm 0.01$ &  $1.00$\\
SF $\gamma=2.1$ & NBC & $+\infty$ & $0.87 \pm 0.01$ &  $1.00$ \\
SF $\gamma=2.5$ & NBC & $+\infty$ & $0.87 \pm 0.01$ & $1.00$ \\
SF $\gamma=3.5$ & NBC & $+\infty$ & $0.89 \pm 0.01$ &  $1.00$ \\
\hline
ER & 2CDC & $+\infty$ &  $0.67 \pm 0.01$ &  $1.00$\\
SF $\gamma=2.1$ & 2CDC & $+\infty$ & $0.64 \pm 0.01$ &  $1.00$ \\
SF $\gamma=2.5$ & 2CDC & $+\infty$ & $0.66 \pm 0.01$ & $1.00$ \\
SF $\gamma=3.5$ & 2CDC & $+\infty$ & $0.67 \pm 0.01$ &  $1.00$ \\
\hline
ER & NBC & $0$ &  $0.35 \pm 0.01$ &  $1.00$\\
SF $\gamma=2.1$ & NBC & $0$ & $0.16 \pm 0.01$ &  $0.99$ \\
SF $\gamma=2.5$ & NBC & $0$ & $0.15 \pm 0.01$ & $0.99$ \\
SF $\gamma=3.5$ & NBC & $0$ & $0.25 \pm 0.01$ &  $1.00$ \\
\hline
\end{tabular}
\caption{{\bf Critical exponents of the network dismantling transition.}
This data refers to the best fits of the scaling of Eq.~(\ref{eq:alpha}) performed in Figure~\ref{fig:4}. Each row in the table identifies a specific network model and a specific percolation process. For each of them, we report the best estimate and associated uncertainty of the exponent $\alpha$ as well as the value of the square of the Pearson correlation coefficient $R^2$. }
\label{tab:exponents_tree}
\end{table}

\subsection{Real-world networks}

One the greatest advantages that Eq.~(\ref{eq:hyperscaling}) provides is enabling a FSS-like analysis for real-world networks. This is far from being a trivial task, as there is no obvious generative model for real networks that would allow us to scale up/down their size, even more so considering that we analyze a corpus including a large variety of real-world networks, displaying marked differences in their topological properties. In Fig.~\ref{fig:5}, we test the scaling of Eq.~(\ref{eq:alpha}) on real networks: each point represents one network in our corpus; for each of them, we identify the pseudo-critical tree emerging under max-NBC-biased percolation, then plot its diameter {\it vs.} its size.  Strikingly, we recover a scaling very similar to the one valid for synthetic graphs. Such a power-law scaling is not as good as observed in Fig.~\ref{fig:4} for synthetic graphs; however, most of the points are still compatible with Eq.~(\ref{eq:alpha}) with $\alpha^{\mathrm{(NBC,2C)}} \simeq 0.88$, i.e., a value of the exponent compatible with what observed on synthetic graphs. This finding reinforces our claim that the max-NBC-biased percolation process washes away any peculiar topological properties of the graph before the actual transition takes place. 
An analogous result is valid also for max-2CDC-biased percolation, where $\alpha^{\mathrm{(2CDC,2C)}} \simeq 0.65$ appears to capture the overall trend well, see Fig.~\ref{fig:5}(b). 
Further, as displayed already by synthetic graphs, the universality of the power-law exponent $\alpha^{\mathrm{(x,2C)}}$ does not hold for non-maximally-biased percolation processes, see Fig.~\ref{fig:5}(c). For real networks, this means that there is not a clear power-law trend as networks differ substantially in their topological properties, including their degree distribution.

\section{Conclusions}

In this paper, we studied the critical properties of the network dismantling transition. We did it by first devising a biased adaptive percolation process that is based on the edge non-backtracking centrality (NBC), which we shown to be a greedy-optimal protocol to dismantle a network.
% reduce the robustness of a network. 
Then, we studied systematically such a NBC-biased percolation process on a large variety of networks. Finite-size scaling (FSS) analysis on random graphs revealed that the transition displayed by the max-NBC strategy is characterized by properties that are identical in both graphs with Poisson and power-law degree distributions. In the transition, the giant component (GC) disappears in a discontinuous manner; the largest 2-core (2C) of the graph vanishes in a continuous manner, but with a critical exponent that appears to be degree-distribution independent. 
Simulations performed on real networks indicate that such an universal behavior of the optimal percolation transition may be robust to variations of several network topological properties, not just the degree distribution. 

Due to the high computational cost of iteratively estimating the NBC metric, we devised a surrogate version named 2-core-degree-centrality(2CDC)-biased percolation. Such a process is less effective than NBC-biased percolation in reducing the robustness of a network; however, it enables the analysis of much larger networks thanks to its lower computational cost. Also, it replicates the same qualitative results of max-NBC-biased percolation, with abrupt GC and 2C transitions that are essentially topology independent.

Our findings suggest that the network dismantling transition belongs to a robust universality class, with critical properties largely independent of the specific network topology. This contrasts sharply with ordinary percolation, where critical behavior can depend strongly on network structure~\cite{cirigliano2024scaling}. These results point toward the possibility of a relatively simple, topology-independent theory describing the critical behavior of network dismantling. Developing such a theory is an interesting direction for future work.

From the complementary perspective of practitioners interested in strategies to prevent network collapse, our results reveal a potential challenge. In spite of being quasi-optimal in dismantling the network, the max-NBC strategy inflicts almost no damage to the GC or the 2C until their abrupt collapse. Consequently, when link removal is guided by the max-NBC strategy, the robustness of the network is silently eroded while these macroscopic observables remain essentially unchanged. Monitoring the size of the GC \cite{mugisha2016identifying}, or even of the 2C, does not provide an early-warning signal of global collapse, and new indicators are needed.

\section*{Code availability}
Code implementing the biased percolation processes studied in this paper can be found at \url{https://github.com/hanlinsun97/criticality-network-dismantling}.

\acknowledgments

F.R.  was partially supported by the Air Force Office of Scientific Research under grant number FA9550-24-1-0039. The funders had no role in study design, data collection, and analysis, the decision to publish, or any opinions, findings, conclusions, or recommendations expressed in the manuscript. 
H.S. was supported by the Wallenberg Initiative on Networks and Quantum Information (WINQ) and by the European Union under the Marie Skłodowska-Curie Actions grant agreement No. 101211574. H.S. is also thankful to J. Liu for insightful discussion on this work.

\appendix

\section{Notation}
\label{app:notation}

In Tab.~\ref{tab:notation}, we summarize the notation used in the paper.

\begin{table*}[!htb]
\begin{tabular}{|c|r|l|}
\hline
context & notation & object represented
\\
\hline \hline
\multirow{9}{*}{\makecell[c]{percolation\\ processes}} & $G = (\mathcal{N}, \mathcal{E})$ & graph composed of a set of nodes and edges
\\ \cline{2-3}
& $\mathcal{N}$, $N$& set of nodes and its size
\\ \cline{2-3}
& $\mathcal{E}$, $E$ & set of edges and its size
\\ \cline{2-3}
& $\vec{\epsilon} = \left(\epsilon_1, \ldots, \epsilon_E \right)$ & \makecell[l]{list of edges that defines a 
specific instance
\\ of a bond-percolation process} 
\\ \cline{2-3}
& $G_t$ & \makecell[l]{remnant graph after the removal of $t$ edges} 
\\ \cline{2-3}
& $s(e, G)$ & score of edge $e$ in graph $G$
\\ \cline{2-3}
& $P(\epsilon_t = e)$ & \makecell[l]{probability of selecting edge $e$\\ at stage $t$ of a biased percolation process}
\\\cline{2-3}
& $S(G)$ & relative size of the GC of graph $G$ 
\\ \cline{2-3}
& $C(G)$ & relative size of the 2C of graph $G$ 
\\ \hline \hline
\multirow{13}{*}{\makecell[c]{numerical\\ simulations}} &
$_qG_t^{(\mathrm{x})}$ & \makecell[l]{remnant graph
after the removal of $t$
edges
\\
in the $q$th instance of 
process x = ORD, NBC, 2CDC}
\\ \cline{2-3}
&
$\langle S^{(\mathrm{x})}(t, G) \rangle = \frac{1}{Q} \sum_{q=1}^Q \, \, S(\, _qG_t^{(\mathrm{x})})$ & \makecell[l]{average value of the GC in $Q$
independent realizations
\\
of process x on graph $G$;
in FSS analysis, we replace $G$ with\\ $N$ to highlight the dependence 
on the network size}
\\ \cline{2-3}
&
$\langle C^{(\mathrm{x})}(t, G) \rangle = \frac{1}{Q} \sum_{q=1}^Q \, \, C(\, _qG_t^{(\mathrm{x})})$ & \makecell[l]{average value of the 2C in $Q$
independent realizations
\\
of process x on graph $G$}
\\ \cline{2-3}
& $_q\hat{t}^{(\mathrm{x, o})}(G)$ & \makecell[l]{pseudo-critical point in instance $q$ as defined in Eq.~(\ref{eq:pseudo_critical});
\\here, o = GC, 2C
}
\\ \cline{2-3}
& $_{q}\hat{p}^{(\mathrm{x, o})}(G)$ &  \makecell[l]{same as above, but quantified in terms of fraction of edges}
\\ \cline{2-3}
& $\langle \hat{p}^{(\mathrm{x, o})}(G) \rangle = \frac{1}{Q} \, \sum_{q=1}^Q \, _{q}\hat{p}^{(\mathrm{x, o})}(G)$ &  \makecell[l]{average value of the pseudo-critical point}
\\ \cline{2-3}
&  $p_c^{\mathrm{(x, o)}} $ &  \makecell[l]{critical threshold of process $x$ for order parameter $o$}
\\ \cline{2-3}
& $\delta^{\mathrm{(x, o)}} (G)  = \left| \langle \hat{p}^{(\mathrm{x, o})}(G) \rangle - p_c^{\mathrm{(x, o)}} \right|$ &  \makecell[l]{distance of the average value of the pseudo-critical point\\ from the critical point}
\\ \cline{2-3}
& $\sigma^{\mathrm{(x, o)}}(G) = \sqrt{  \frac{1}{Q} \, \sum_{q=1}^Q \left[ \, _{q}\hat{p}^{(\mathrm{x, o})}(G) \right]^2 - \left[ \langle \hat{p}^{(\mathrm{x, o})}(G) \rangle \right]^2}$ &  \makecell[l]{standard deviation of the pseudo-critical point}
\\ \cline{2-3}
& $_{q}\hat{S}^{\mathrm{(x, o)}}(G)$, $_{q}\hat{C}^{\mathrm{(x, o)}}(G)$ & \makecell[l]{values of the order parameters at the
pseudo-critical point\\in instance $q$ of the process $x$ applied to graph $G$, \\
when pseudo-criticality is based on order parameter o}
\\ \cline{2-3}
& $\langle \hat{S}^{\mathrm{(x, o)}}(G) \rangle$, $\langle \hat{C}^{\mathrm{(x, o)}}(G) \rangle$ & \makecell[l]{average values of the above quantities}
\\ \cline{2-3}
& $\beta^{(\mathrm{x, o})}$, $\bar{\nu}^{(\mathrm{x, o})}_{1}$, $\bar{\nu}^{(\mathrm{x, o})}_{2}$ & \makecell[l]{critical exponents for the transition of order parameter o \\
in process $\mathrm{x}$}
\\ \hline \hline
\multirow{5}{*}{NB theory} & $\overline{\mathcal{E}}$ & set of directed edges
\\ \cline{2-3}
& $B(G)$ & NB operator of graph $G$
\\ \cline{2-3}
& $\Lambda(G)$ & principal NB eigenvalue of graph $G$
\\ \cline{2-3}
& $\ket{r}(G)$ & principal NB right eigenvector of graph $G$ 
\\ \cline{2-3}
& $\bra{\ell}(G)$ & principal NB left eigenvector of graph $G$ 
\\ \hline 
\end{tabular}
\caption{{\bf List of symbols used in the paper.} The symbols are grouped according to the part of the paper in which they are introduced, and a brief description of their meaning is provided. In the table, we use the following abbreviations: GC = giant component, 2C = 2-core, NB = nonbracktracking, FSS = finite-size scaling, ORD = ordinary, NBC = non-backtracking centrality, 2CDC = 2-core degree centrality.}
\label{tab:notation}
\end{table*}

\section{Non-backtracking centrality}
\label{appendix:perturbation}

Let us choose the vectors $\ket{i \tight{\to} j}$ as a complete, orthogonal basis for the $2E$-dimensional space of the directed edges $\overline{\mathcal{E}}$ of a graph $G$, such that $\sum_{i \to j \in \overline{\mathcal{E}}} \dyad{i \tight{\to} j}{i \tight{\to} j} = \mathbb{1}$ is the identity, and $\braket{i \tight{\to} j}{k \tight{\to} l} = \delta_{i \to j, k \to l}=\delta_{i,k} \delta_{j,l}$. The elements of the NB matrix in this basis are given by~Eq.\eqref{eq:matrix_elements_B}.

We denote with $\Lambda$ the leading NB eigenvalue. By the Perron-Frobenius theorem, $\Lambda \in \mathbb{R}$; further, $\Lambda>1$ if the network contains at least two connected cycles~\cite{torres2020non}. The left $\ket{\ell}$ and right $\ket{r}$ eigenvectors associated to $\Lambda$, defined by $\bra{\ell}B=\Lambda \bra{\ell}$ and $B\ket{r}=\Lambda \ket{r}$, respectively, can be chosen with non-negative entries and normalized as $\braket{\ell}{r}=1$.

Suppose we remove the edge $(x,y)$ from $G$. This is a rank-2 perturbation on $B$ that sets equal to zero the entries in $B$ corresponding to the basis vectors $\ket{x \tight{\to} y}$  and
$\ket{y \tight{\to} x}$. We can write such a perturbation as
\begin{equation}
    \delta B = (\mathbb{1}-Q)B(\mathbb{1}-Q)-B=-(QB+BQ),
    \label{eq:delta_B}
\end{equation}
where $Q=\dyad{x \tight{\to} y}{x \tight{\to} y}+\dyad{y \tight{\to} x}{y \tight{\to} x}$, and the term $QBQ=0$ by definition of $B$. Assuming that the principal eigenvector is not affected by the perturbation, the perturbed eigenvector equation reads
\begin{equation}
    (B+\delta B) \ket{r} \simeq (\Lambda + \delta \Lambda) (\mathbb{1}-Q) \ket{r} .
    \label{eq:perturbation}
\end{equation}
Multiplying both sides of the above equation by $\bra{\ell}$, we get $
\Lambda + \bra{\ell}\delta B \ket{r} = \Lambda + \delta \Lambda - (\Lambda + \delta \Lambda) \bra{\ell}Q\ket{r}$, which simplifies to
 $
\bra{\ell}\delta B \ket{r} = \delta \Lambda - (\Lambda + \delta \Lambda) \bra{\ell}Q\ket{r}$.
Using Eq.~(\ref{eq:delta_B}), we have that $\bra{\ell}\delta B \ket{r} = - 2 \Lambda \bra{\ell}Q\ket{r}$, thus the previous equation becomes $
- \Lambda \bra{\ell} Q \ket{r} = \delta \Lambda - \delta \Lambda \bra{\ell}Q\ket{r}$, from which it follows that
\begin{equation}
    \delta \Lambda = - \Lambda \, \frac{\bra{\ell}Q\ket{r}}{1 - \bra{\ell}Q\ket{r}} \; .
\end{equation}
Since $\sum_{(x,y) \in \mathcal{E}}Q = \sum_{(x,y) \in \mathcal{E}} \dyad{x \tight{\to} y}{x \tight{\to} y}+\dyad{y \tight{\to} x}{y \tight{\to} x} = \mathbb{1}$, we have $\sum_{(x,y) \in \mathcal{E}} \bra{\ell}Q\ket{r}=2E$. Thus if $\ket{r}$ is not localized, $\bra{\ell}Q\ket{r} \ll 1$, and we can approximate at first order
\begin{equation}
    \delta \Lambda \simeq - \Lambda \bra{\ell}Q\ket{r} \; .
    \label{eq:del}
\end{equation}
Expanding the left and right eigenvectors in the canonical basis $\ket{i \tight{\to} j}$, we just have to compute the matrix elements 
\begin{equation*}
    \braket{i \tight{\to} j}{Q|k \tight{\to} l} = \delta_{i \to j, x\to y}\delta_{k \to l, x \to y} + \delta_{i \to j, y\to x}\delta_{k \to l, y \to x},
\end{equation*}
from which we finally get 
Eq.~\eqref{eq:var_lambda} in the main text. 

\begin{figure}
    \centering
    \includegraphics[width=0.95\linewidth]{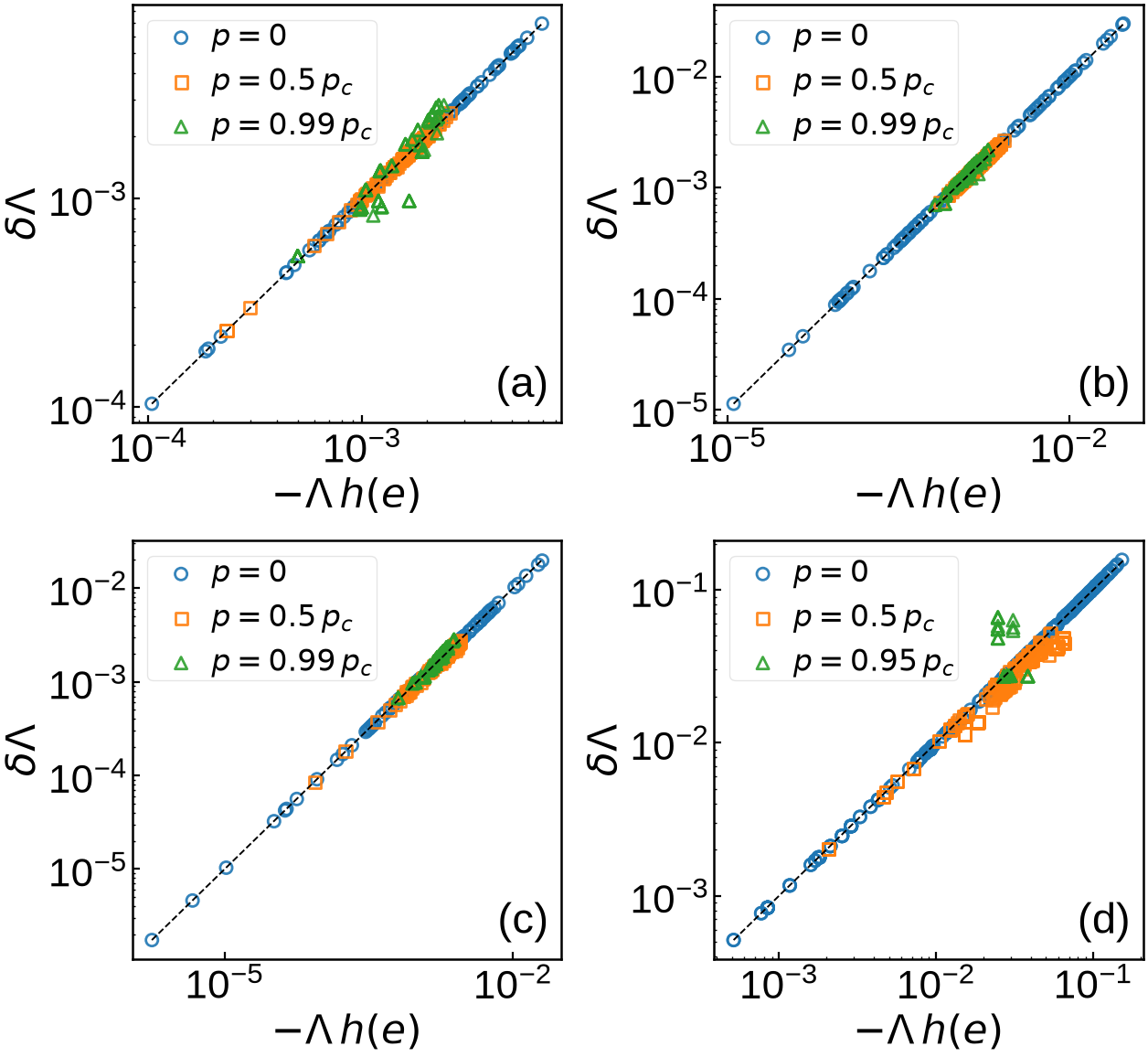}
    \caption{{\bf NBC centrality and eigenvalue perturbation.} 
    Numerical comparison between $-\Lambda h(e)$ -- the r.h.s of Eq.~\eqref{eq:var_lambda} -- and the actual effect of an edge removal on the spectral radius, $\delta \Lambda$ -- the l.h.s of Eq.~\eqref{eq:var_lambda}. The two quantities are computed numerically at different steps $t=pE$ of a max-NBC percolation process, different symbols denoting different fractions $p<p_c=1-(N-n_C)/E$ of edges removed. Various networks have been considered: (a) ER networks with $\langle k \rangle=4$, (b) scale-free networks with $k_{\text{min}}=3$, $\gamma=2.5$, (c) clustered random networks with $\langle k \rangle = 6$ and global clustering coefficient $T=0.38$, generated from an ER backbone following Ref.~\cite{cirigliano2024strongly}, (d) ``Les Miserables'' real network~\cite{knuth1993the}. Synthetic networks in (a)-(c) have $N=1000$, while $N=77$, $E=254$ in (d). Apart from small deviations very close to $p_c$, Eq.~\eqref{eq:var_lambda} holds for all networks considered.}
    \label{fig:empirical_delta}
\end{figure}

Some remarks are in order here. First, we have neglected the effect of the removal of the edge $e$ on the eigenvectors. While this is usually a safe assumption if we are far from the percolation threshold, when we approach the transition and $\Lambda \tight{\to} 1$ the contribution coming from the perturbed eigenvectors can be relevant. This means that the shift $\delta \Lambda$ may no longer be simply expressed as in Eq.~\eqref{eq:var_lambda}. Second, we have made no assumptions on the structure of $G$, apart from it being connected and with at least two loops. It is thus interesting to see how well Eq.~\eqref{eq:var_lambda} works depending on the properties of the graph $G$. For this reason, we set up the following numerical experiment. We perform a max-NBC percolation process on a graph $G$. When some fraction $p<p_c=1-(N-n_C)/E$ of the edges has been removed, we measure numerically $\Lambda(G_{t=pE})$, the NBC $h(e,G_{t=pE})$ of some remaining edges, the true $\delta \Lambda$ that the removal of such edges would produce, in order to compare the two sides of Eq.~\eqref{eq:var_lambda}. Then we keep removing edges with the max-NBC greedy strategy, starting from the graph $G_{t=pE}$. Results of this experiment for (a) ER networks, (b) SF networks, (c) strongly clustered random networks, and (d) real-world networks, are reported in Figure~\ref{fig:empirical_delta}, showing that Eq.~\eqref{eq:var_lambda} works well for all the networks considered, and small discrepancies are observed only very close to $p_c$.

%%%%%%%%%%%%%%%%%%%%%%%

\section{Non-backtracking centrality and participation to loops}
\label{appendix:NBC_loops}

Here, we discuss the relation between the NBC of an edge $e$ -- a centrality measure derived from the spectral properties of the NB matrix -- and the number of cycles to which $e$ belongs. 

\begin{figure}
    \centering
    \includegraphics[width=0.95\linewidth]{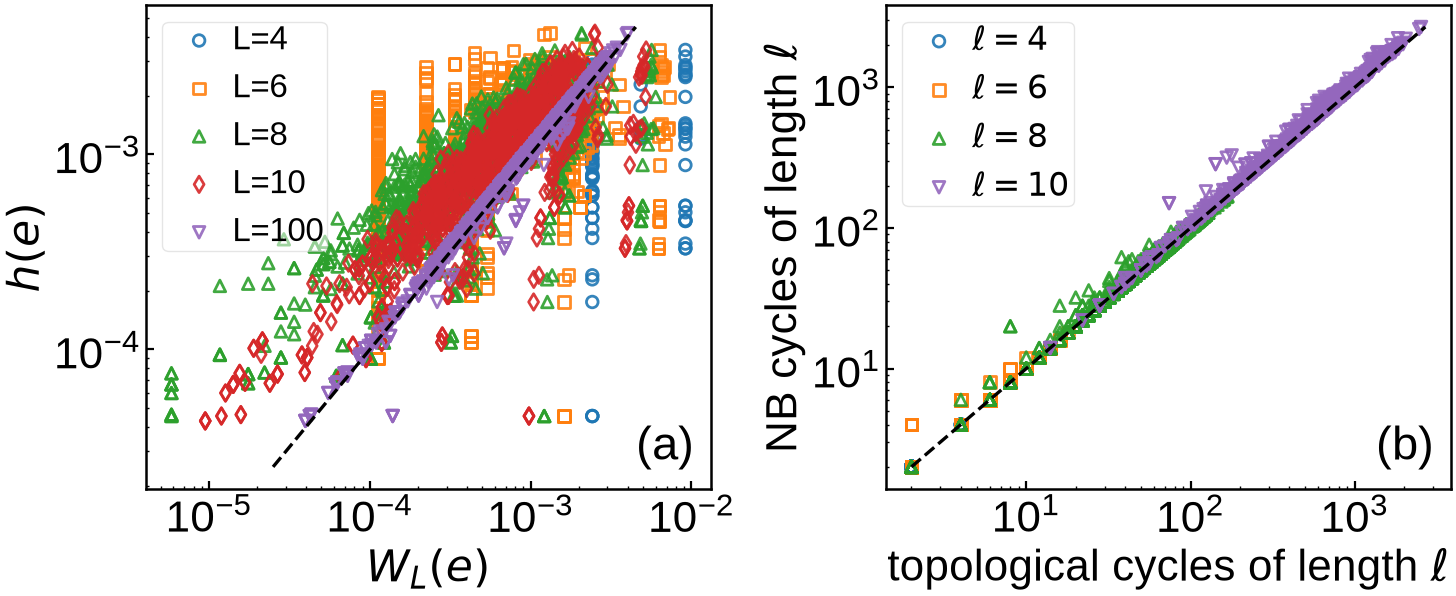}
    \caption{{\bf NBC and loops density.} 
    (a) NBC vs the r.h.s. of Eq.\eqref{eq:h_vs_loops}, here denoted with $W_L(e)$, for various values of $L$. The asymptotic relation expressed in Eq.~\eqref{eq:h_vs_loops} is more accurate as $L$ increases. (b) The relation between the number of topological cycles and the number of NB cycles of length $\ell$. Dashed lines have slope $1$. A small ER network of size $N=512$ and $c=4$ has been used for both panels.}
    \label{fig:empirical_loops}
\end{figure}

Let us first define the spectral decomposition of $B$ in terms of its left and right eigenvectors. Let $\{\lambda_1=\Lambda, \lambda_2, \ldots, \lambda_{2E}\}$ be the eigenvalues of $B$, and let us denote with $\ket{u^n}$ and $\ket{v^{n}}$ the left and right eigenvectors, respectively, associated to the eigenvalue $\lambda_n$. 
We can then write
\begin{equation}
    B = \sum_{n} \lambda_n \dyad{u^n}{v^n} = \Lambda \dyad{r}{\ell} + \sum_{n \geq 2} \lambda_n \dyad{u^n}{v^n}.
    \label{eq:spectral_decomposition}
\end{equation}
For a directed edge $i \tight{\to} j$, let us introduce the edge-local resolvent 
\begin{equation}
	H_{i \to j}(z) = \sum_{k=1}^{\infty}(B^k)_{i \to j,i \to j} z^k.
\end{equation}
This is the generating function of the number of NB loops the directed edge $i \tight{\to} j$ belongs to. Using the spectral decomposition Eq.~\eqref{eq:spectral_decomposition} and summing the geometric series, we get
\begin{equation}
	H_{i \to j}(z)=\sum_{n} \frac{u^n_{i \to j}v^n_{i \to j}}{1-\lambda_n z},
\end{equation}
where $|z| < 1/\Lambda$. The leading singularity of $H_{i \to j}(z)$ is at $z = 1/\Lambda$. Thus we have for $z \uparrow 1/\Lambda$
\begin{equation}
	H_{i \to j}(z) \sim r_{i \to j} \ell_{i \to j}  (1-\Lambda z)^{-1}.
\end{equation}
Now we can use the Hardy-Littlewood Tauberian theorem~\cite{titchmarsh1939theory}, which states that if $a_n \geq 0$ for all $n \in \mathbb{N}$, and $\sum_{k=0}^{\infty} a_k x^k \sim \frac{1}{1-x}$, 
then we have $\sum_{k=0}^{L} a_k \sim L$ as $L \to \infty$.
Applying this theorem for the series with coefficients $a_0=0$ and $a_k = \frac{(B^{k})_{i \to j, i \to j}}{\Lambda^k r_{i \to j} \ell_{i \to j}}$,
we get asymptotically for large $L$
\begin{equation}
	\label{eq:cycles_asymptotics}
	r_{i \to j} \ell_{i \to j} \sim \frac{1}{L} \sum_{k=0}^{L} \frac{(B^{k})_{i \to j, i \to j}}{\Lambda^k}.
\end{equation}
Taking $H_{i \to j}+H_{j \to i}$, and using the fact that for undirected graphs a directed edge and its reversed belong to the same cycles, we get 
\begin{equation}
    h(e) \sim  \frac{2}{L} \sum_{k=0}^{L} \frac{(B^{k})_{i \to j, i \to j}}{\Lambda^k}.
    \label{eq:h_vs_loops}
\end{equation}
Remarkably, Eq.~\eqref{eq:h_vs_loops} is telling us that the more NB loops an edge belongs to, the higher the value of its NBC weight~\footnote{In principle, Eq.~\eqref{eq:h_vs_loops} counts all NB cycles, even those obtained from multiple repetitions of shorter NB cycles. For instance, a triangle traversed twice counts as a loop of length $6$. In practice, however, such cycles are exponentially suppressed by the factor $\Lambda^{-k}$. In fact, the number of triangles and of $6$-loops made of double traversal of triangles is the same. But the latter gets an extra factor $\Lambda^{3}$ in the denominator.}. Note that an edge that belongs to many short loops will typically also belong to long loops, while the viceversa is not true. Thus edges with high $h(e,G)$ are in the denser, more connected region of the 2C of $G$.
In Figure~\ref{fig:empirical_loops} we check numerically Eq.~\eqref{eq:h_vs_loops}. For all edges of small ER networks of size $N=1000$, we compute their NBC, the number of NB cycles and the number of primitive loops of various length they belong to: panel (a) shows that Eq.~\eqref{eq:h_vs_loops} becomes more precise as $L$ increases, with a perfect correlation already for $L \simeq 10$; panel (b) shows that the number of NB loops and of topological loops of length $\ell$, both weighted with $\Lambda^{-\ell}$ to suppress non-primitive cycles, are strongly correlated. Thus we can conclude that the NBC ranks edges according to their participation to topological loops, of all lengths.

\section{Percolation threshold for the dismantling of uncorrelated random graphs}
\label{appendix:threshold}

On a finite network $G$ with $N$ nodes and $E$ edges, Eq.~\eqref{eq:feedback_edge_set} gives the fraction of edges that the network have in excess to a loop-less structure, i.e., a forest. In the limit $N \to \infty$, Eq.~\eqref{eq:feedback_edge_set} defines the percolation threshold for both NBC- and 2CDC-biased percolation. For this purpose, one needs to known in advance how the number of edges $E$ and the number of connected components $n_C$ of the graph scale with $N$, and then take the limit $N \to \infty$.

While studying random graphs with arbitrary degree distributions $P(k)$, the number of nodes $N$ is given. However, $E$ and $n_C$ are generally random variates obeying specific distributions whose functional form depends non-trivially on $P(k)$. In this section, we derive an expression for Eq.~\eqref{eq:feedback_edge_set} on the ensemble of uncorrelated random graphs with arbitrary degree distribution $P(k)$. 
Given the number of nodes $N$, averaging over the ensemble we have $E= \langle k \rangle N/2$. For the number of connected components $n_C$, Eq.~(55) in~\cite{cirigliano2024scaling} gives $n_C = N F+1$, where $F = g_0(z)-\frac{\langle k \rangle}{2} z^2$,
$g_0(x) = \sum_{k} P(k)x^k$ is the generating function of the degree distribution, and $z$ denotes the probability that a randomly chosen link does not lead to the giant component, which satisfies Eq.~\eqref{eq:recursive}.
Note that here we assume the finite clusters are tree-like and include a negligible number of cycles. Thus, we have $N-n_C \simeq N(1-F)$, 
from which we obtain Eq.~\eqref{eq:threshold}.
Such an
expression serves to estimate the thermodynamic critical point of the biased percolation processes considered in the paper. The expression is used when studying the FSS scaling of Eq.~(\ref{eq:scaling_pseudo1}).
Note that if the random graph has a single giant connected component, i.e., $n_C=1$, Eq. \eqref{eq:feedback_edge_set}
can be greatly simplified as
\begin{equation}
p_c^{\mathrm{(x,o)}} 
= 1 - \frac{2}{\av{k}}.
\label{eq:threshold_single_component}
\end{equation}
We can now use Eq.\eqref{eq:threshold} for specific cases by simply plugging the desired degree distribution $P(k)$. For example, for ER networks with infinite size, we should use the Poisson distribution $P(k) = \frac{e^{-\langle k \rangle} \, \langle k \rangle^k}{k!}$. For the specific case $\av{k}=4$ considered in the main text, we get $p_c^{\mathrm{(x,o)}}=0.5095205748\ldots$
For the case of scale-free networks considered in the main text, where $k_{\min}=3$ and the entire network belongs to a giant 2-core hence $n_C = 1$, we can use instead Eq.~\eqref{eq:threshold_single_component}.

\section{max-NBC-biased percolation on scale-free random graphs}
\label{app:SF}

\begin{figure}[!htb]
    \centering
    \includegraphics[width=0.45\textwidth]{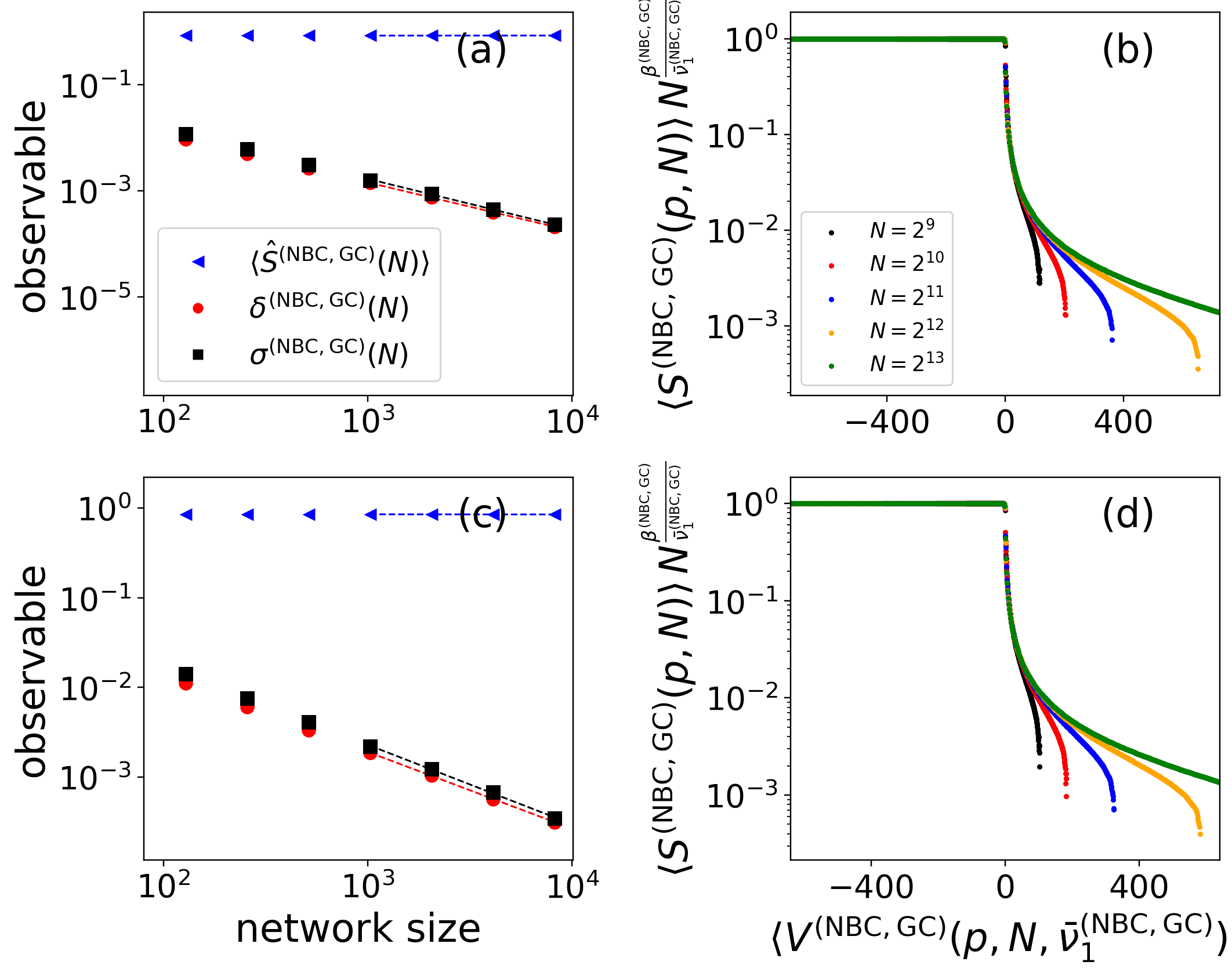}
    \caption{{\bf Critical properties of the GC transition for NBC-biased percolation on scale-free random graphs.}  (a) Same as in Fig.~\ref{fig:2}(c), but for the NBC-biased percolation process with $a = + \infty$ applied to random graphs with power-law degree distribution, i.e., $P(k) \sim k^{-\gamma}$ if $ k \in [k_{\min}, k_{\max}] $ and $P(k) = 0 $ otherwise, generated according to the configuration model. Here, we set $\gamma=2.5$, $k_{\min}=3$, and $k_{\max}=\sqrt{N}$. 
    (c-d) Same as in (a) and (b), respectively, but for scale-free graphs with degree exponent $\gamma = 3.5$.  
    Results are averaged over $Q = 5000$ instances of the biased percolation model.
    Best estimates of the critical exponents are reported in Tab.~\ref{tab:exponents}.
    }
    \label{fig:4old}
\end{figure}

\begin{figure}[!htb]
    \centering
    \includegraphics[width=0.45\textwidth]{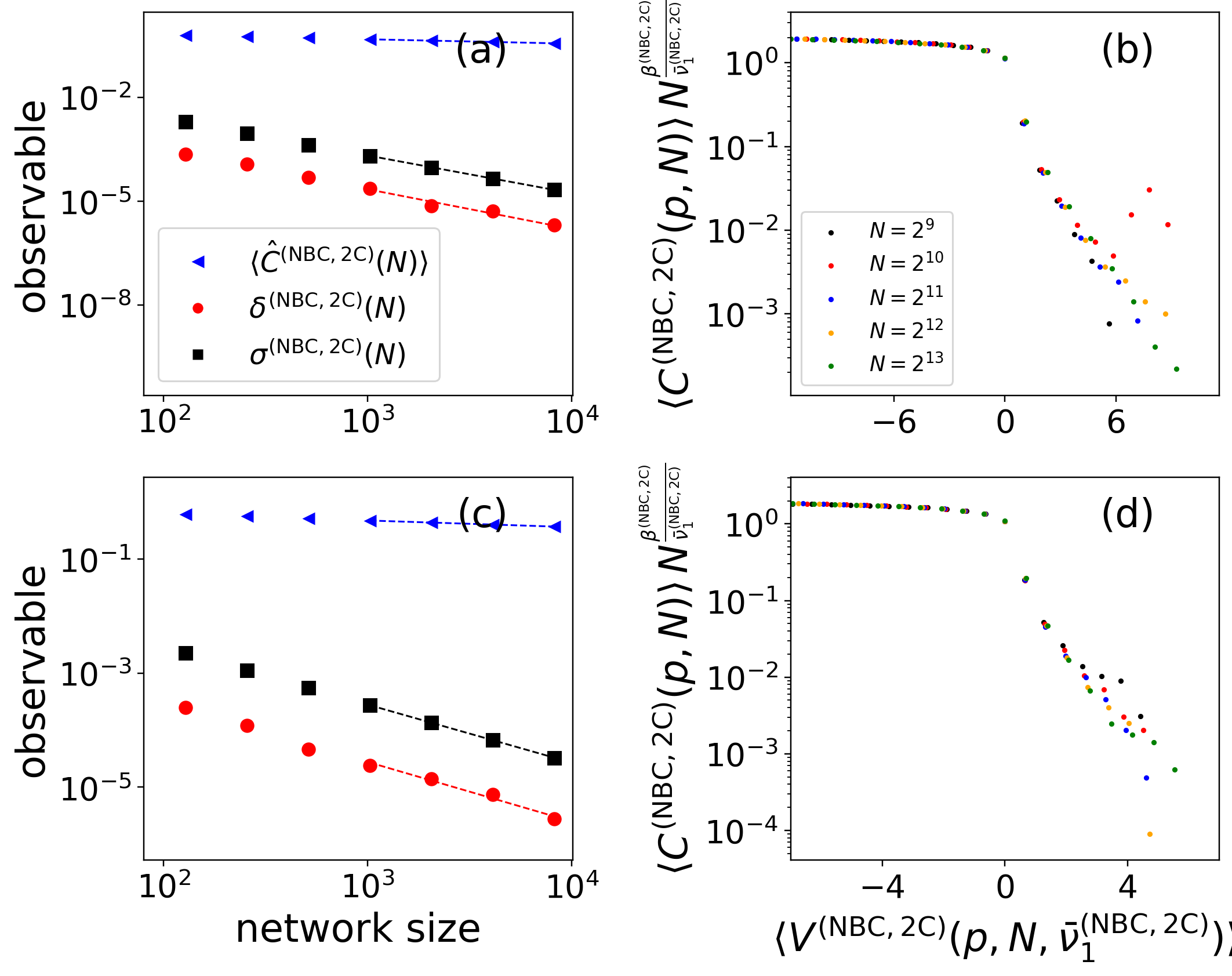}
    \caption{{\bf Critical properties of the 2C transition for NBC-biased percolation on scale-free random graphs.}  (a) Same as in Fig.~\ref{fig:3}(a), but for the NBC-biased percolation process with $a = + \infty$ applied to random graphs with power-law degree distribution, i.e., $P(k) \sim k^{-\gamma}$ if $ k \in [k_{\min}, k_{\max}] $ and $P(k) = 0 $ otherwise, generated according to the configuration model. Here, we set $\gamma=2.5$, $k_{\min}=3$, and $k_{\max}=\sqrt{N}$. 
    (b) Same as in Fig.~\ref{fig:3}(d), but for random scale-free graphs. We measure $\beta = 0.10$. (c-d) Same as in (a) and (b), respectively, but for scale-free graphs with degree exponent $\gamma = 3.5$.  
    Results are averaged over $Q = 5000$ instances of the biased percolation model.
    Best estimates of the critical exponents are reported in Tab.~\ref{tab:exponents}.
    }
    \label{fig:old5}
\end{figure}

In Figs.~\ref{fig:4old} and~\ref{fig:old5}, we report results valid for the max-NBC-biased percolation process applied to graphs constructed according to the uncorrelated configuration model~\cite{catanzaro2005generation}. An instance of the model with size $N$ is obtained by first generating a degree sequence consisting of $N$ random integer variables extracted from the power-law distribution $P(k) \sim k^{-\gamma}$ if $k \in [3, \sqrt{N}]$, and  $P(k) = 0$ otherwise. After that, edges are created at random by preserving such a degree sequence.

\bibliography{nbc.bib}

%%%%%%%%%%%%%%%%%%%%%%%%%%%%%%%%%%%%%%%%%%

\clearpage

\newpage

\onecolumngrid

\section*{Supplemental Material}

\setcounter{page}{1} 
\setcounter{figure}{0} 
\setcounter{equation}{0} 
\setcounter{table}{0} 

\renewcommand{\theequation}{S\arabic{equation}}
\renewcommand{\thefigure}{S\arabic{figure}}
\renewcommand{\thetable}{S\arabic{table}}

\begin{figure}[!htb]
    \centering
    \includegraphics[width=0.45\linewidth]{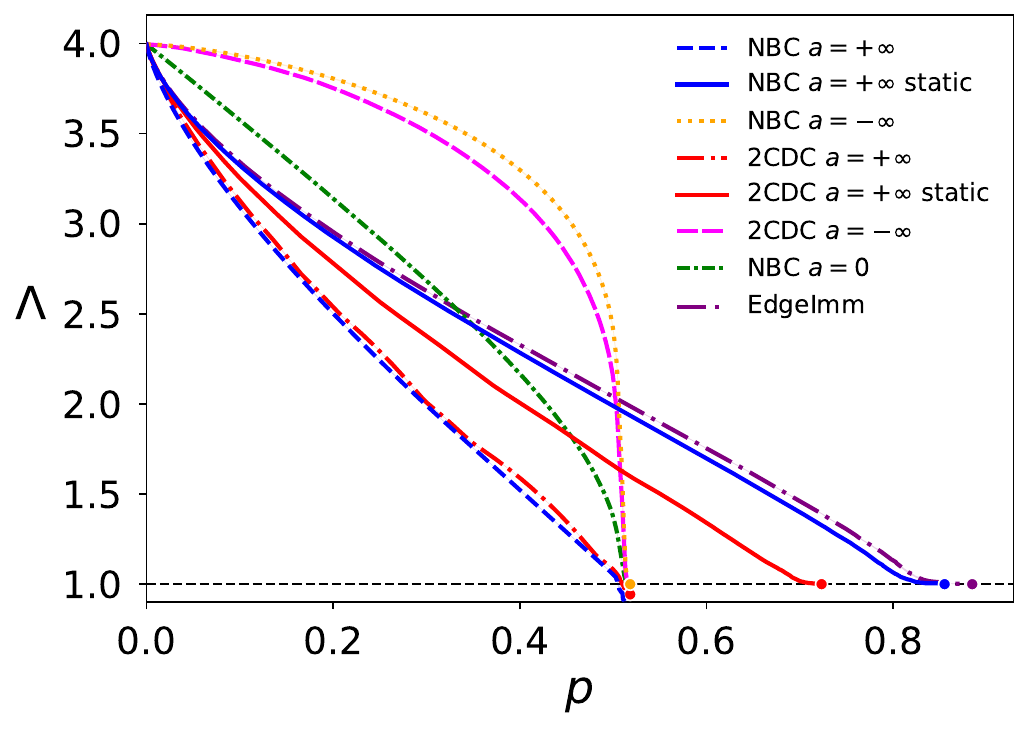}
    \caption{{\bf Quasi-optimality of NBC-biased percolation.} Same as in Fig.~\ref{fig:NBC_vs_others} of the main text.
    We include here some additional curves corresponding to different types of edge-removal strategies.
    }
    \label{fig:NBC_SM}
\end{figure}

\begin{table*}[!htb]
\begin{tabular}{lccccccc}
network model & $x$ & $a$ & OP & $\beta$ & $\bar\nu_1$ & $\bar\nu_2$ & $\alpha$  \\ \hline
%SF $\gamma=2.50$ & NBC & $+\infty$ & GC & 0.00 $\pm$ 0.01 & 1.09 $\pm$ 0.01 & 1.06 $\pm$ 0.01 & 0.87 $\pm$ 0.01  \\ 
%SF $\gamma=3.50$ & NBC & $+\infty$ & GC & 0.00 $\pm$ 0.01 & 1.17 $\pm$ 0.01 & 1.13 $\pm$ 0.01 & 0.88 $\pm$ 0.01 \\ 
%ER & NBC & $+\infty$ & GC & 0.00 $\pm$ 0.01 & 1.23 $\pm$ 0.01 & 1.39 $\pm$ 0.02 & 0.87 $\pm$ 0.01  \\ 
%SF $\gamma=2.50$ & NBC & $+\infty$ & 2C & 0.11 $\pm$ 0.01 & 0.87 $\pm$ 0.03 & 0.92 $\pm$ 0.01 & 0.87 $\pm$ 0.01 \\ 
%SF $\gamma=3.50$ & NBC & $+\infty$ & 2C & 0.12 $\pm$ 0.01 & 0.96 $\pm$ 0.03 & 0.98 $\pm$ 0.01 & 0.88 $\pm$ 0.01 \\ 
%ER & NBC & $+\infty$ & 2C & 0.11 $\pm$ 0.01 & 0.96 $\pm$ 0.02 & 1.98 $\pm$ 0.01 & 0.87 $\pm$ 0.01 \\ 
%\hline
ER & 2CDC & $+\infty$ & GC & 0.00 $\pm$ 0.01 & 1.47 $\pm$ 0.02 & 1.80 $\pm$ 0.01 & 0.67 $\pm$ 0.01\\ 
ER & 2CDC & $0$ & GC & 0.00 $\pm$ 0.01 & 2.87 $\pm$ 0.03 & 2.91 $\pm$ 0.02 & 0.35 $\pm$ 0.01\\ 
ER & 2CDC & $-\infty$ & GC & 0.00 $\pm$ 0.01 & 3.06 $\pm$ 0.05 & 3.36 $\pm$ 0.05 & 0.29 $\pm$ 0.01  \\ 
\hline
ER & 2CDC & $+\infty$ & 2C & 0.33 $\pm$ 0.01 & 1.00 $\pm$ 0.03 & 1.98 $\pm$ 0.01 & 0.67 $\pm$ 0.01  \\ 
ER & 2CDC & $0$ & 2C & 0.67 $\pm$ 0.05 & 1.04 $\pm$ 0.08 & 1.94 $\pm$ 0.01 & 0.35 $\pm$ 0.01  \\ 
ER & 2CDC & $-\infty$ & 2C & 0.71 $\pm$ 0.06 & 1.10 $\pm$ 0.09 & 1.96 $\pm$ 0.01 & 0.29 $\pm$ 0.01  \\ 
\hline
SF $\gamma=2.50$ & 2CDC & $+\infty$ & GC & 0.00 $\pm$ 0.01 & 1.46 $\pm$ 0.01 & 1.46 $\pm$ 0.01 & 0.66 $\pm$ 0.01 \\ 
SF $\gamma=3.50$ & 2CDC & $+\infty$ & GC & 0.00 $\pm$ 0.01 & 1.48 $\pm$ 0.01 & 1.48 $\pm$ 0.01 & 0.67 $\pm$ 0.01 \\ 
SF $\gamma=2.50$ & 2CDC & $0$ & GC & 0.00 $\pm$ 0.01 & 5.58 $\pm$ 0.17 & 6.05 $\pm$ 0.14 & 0.15 $\pm$ 0.01 \\ 
SF $\gamma=3.50$ & 2CDC & $0$ & GC & 0.00 $\pm$ 0.01 & 3.81 $\pm$ 0.06 & 3.90 $\pm$ 0.06 & 0.25 $\pm$ 0.01 \\ 
SF $\gamma=2.50$ & 2CDC & $-\infty$ & GC & 0.00 $\pm$ 0.01 & 16.64 $\pm$ 0.86 & 17.38 $\pm$ 0.88 & 0.05 $\pm$ 0.01  \\ 
SF $\gamma=3.50$ & 2CDC & $-\infty$ & GC & 0.00 $\pm$ 0.01 & 6.35 $\pm$ 0.17 & 7.63 $\pm$ 0.15 & 0.15 $\pm$ 0.01  \\ 
\hline
SF $\gamma=2.50$ & 2CDC & $+\infty$ & 2C & 0.34 $\pm$ 0.01 & 0.98 $\pm$ 0.01 & 0.98 $\pm$ 0.01 & 0.66 $\pm$ 0.01 \\ 
SF $\gamma=3.50$ & 2CDC & $+\infty$ & 2C & 0.33 $\pm$ 0.01 & 1.00 $\pm$ 0.01 & 1.00 $\pm$ 0.01 & 0.67 $\pm$ 0.01  \\ 
SF $\gamma=2.50$ & 2CDC & $0$ & 2C & 0.80 $\pm$ 0.01 & 0.91 $\pm$ 0.01 & 0.96 $\pm$ 0.05 & 0.15 $\pm$ 0.01  \\ 
SF $\gamma=3.50$ & 2CDC & $0$ & 2C & 0.75 $\pm$ 0.01 & 1.00 $\pm$ 0.01 & 1.07 $\pm$ 0.05 & 0.25 $\pm$ 0.01  \\ 
SF $\gamma=2.50$ & 2CDC & $-\infty$ & 2C & 1.01 $\pm$ 0.01 & 1.05 $\pm$ 0.01 & 1.09 $\pm$ 0.01 & 0.05 $\pm$ 0.01  \\ 
SF $\gamma=3.50$ & 2CDC & $-\infty$ & 2C & 0.92 $\pm$ 0.01 & 1.07 $\pm$ 0.01 & 1.19 $\pm$ 0.01 & 0.15 $\pm$ 0.01 \\ 
\hline
\end{tabular}

\caption{{\bf Critical exponents of 2CDC-biased percolation transitions on Erd\H{o}s-R\'enyi and scale-free networks.} Similar to Tab. \ref{tab:exponents}, we list here the best estimates of the critical exponents characterizing the 2CDC-biased percolation process. 
From left to right, we report the network model, the specific order parameter monitored, the bias parameter, and the corresponding values of the critical exponents characterizing the percolation transition. Uncertainties below $0.01$ are rounded up to $0.01$ in the table.  
}
\label{tab:exponents_SF_full}
\end{table*}

 \begin{figure*}[!htb]
     \centering
     \includegraphics[width=0.99\linewidth]{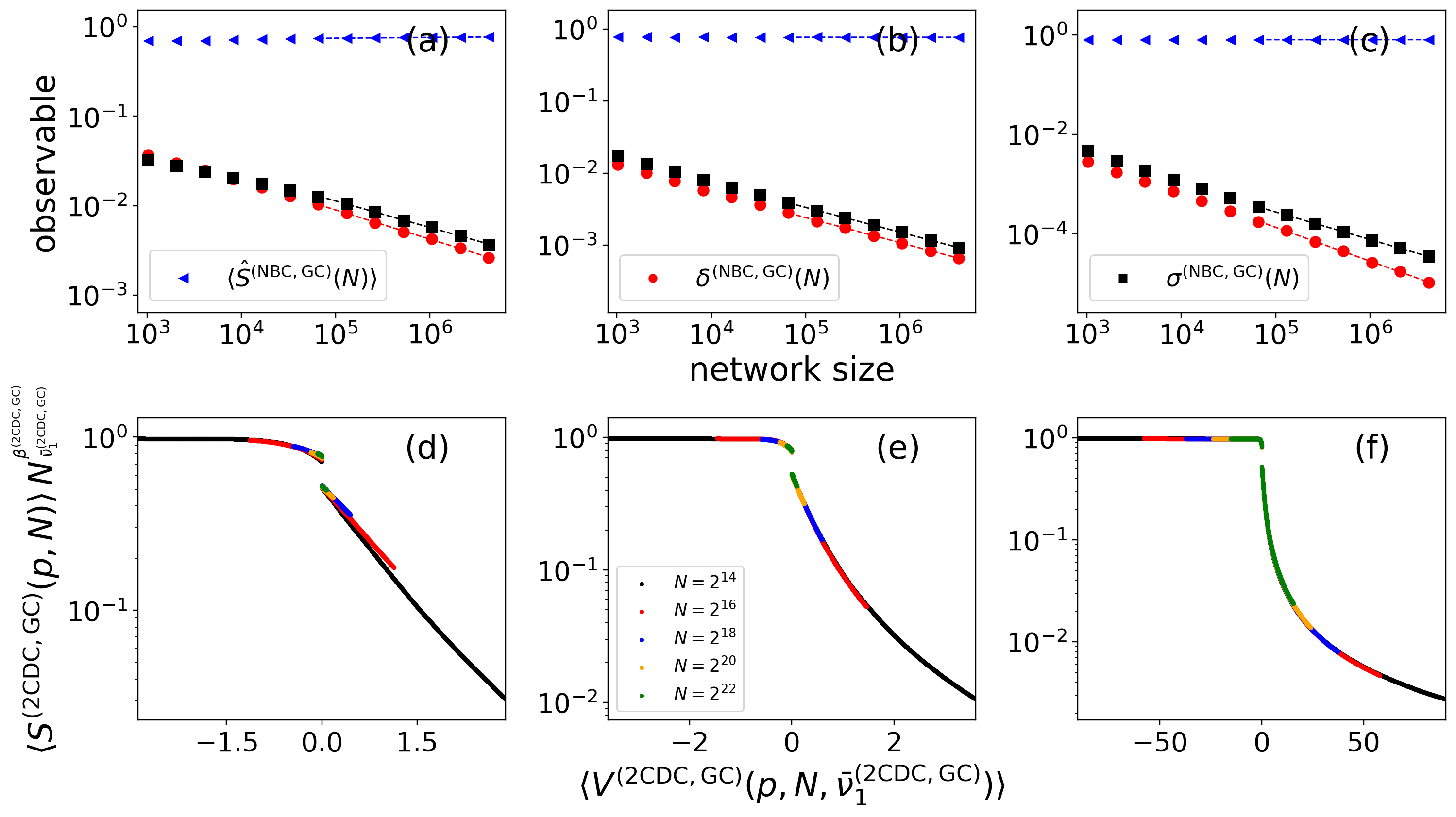}
     \caption{{\bf Critical properties of the GC transition for 2CDC-biased percolation on Erd\H{o}s-R\'enyi graphs.} Same as in Figure~\ref{fig:2}, but for the 2CDC-biased percolation process. 
     Results are averaged over $Q = 10000$ instances of the biased percolation model.
    Best estimates of the critical exponents are reported in Tab.~\ref{tab:exponents_SF_full}.}
     \label{fig:degree_2CDCGC}
 \end{figure*}

 \begin{figure*}[!htb]
     \centering
     \includegraphics[width=0.99\linewidth]{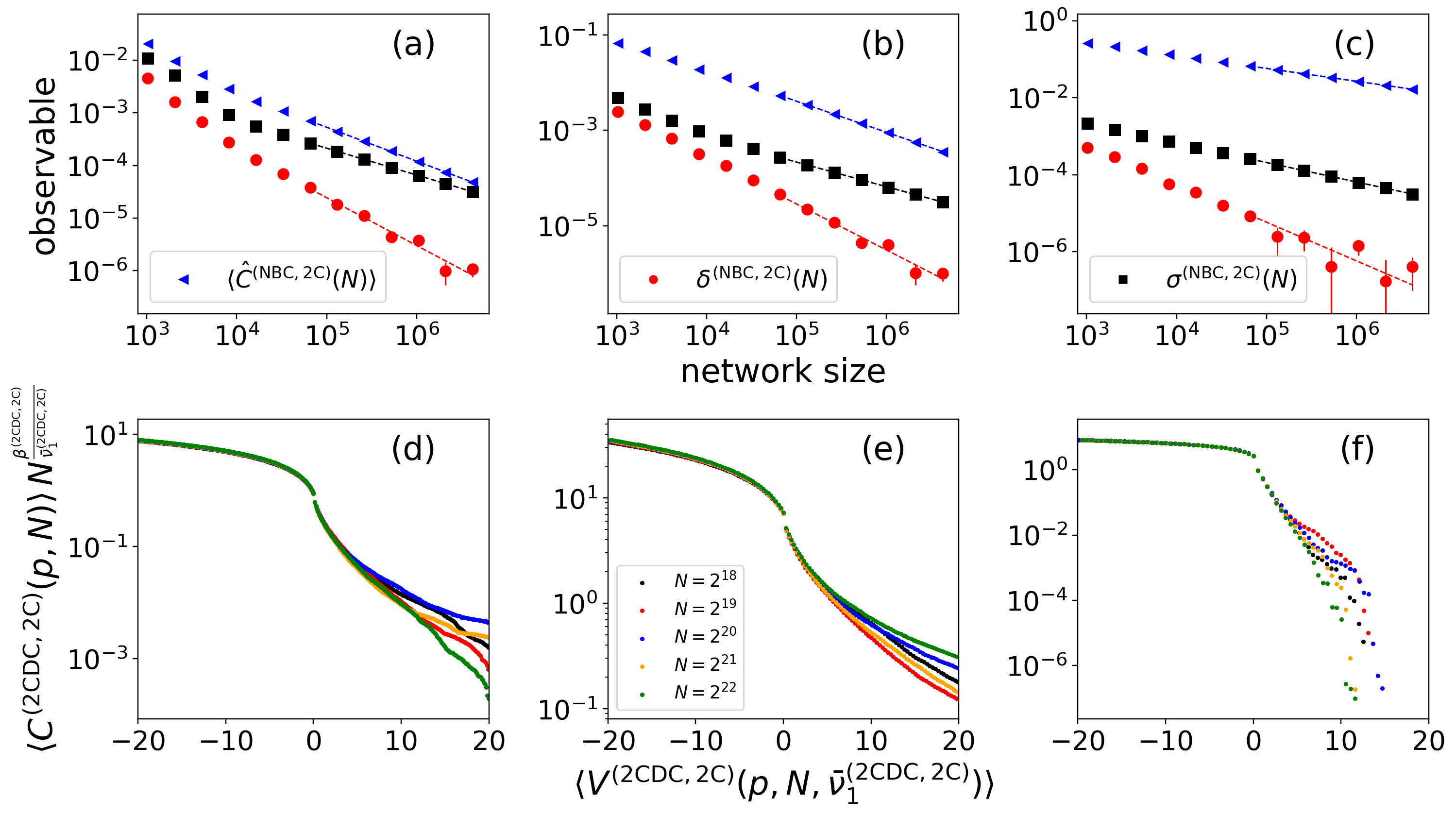}
     \caption{{\bf Critical properties of the 2C transition for 2CDC-biased percolation on Erd\H{o}s-R\'enyi graphs.} 
     Same as in Figure~\ref{fig:3}, but for the 2CDC-biased percolation process. 
     Results are averaged over $Q = 10000$ instances of the biased percolation model.
    Best estimates of the critical exponents are reported in Tab.~\ref{tab:exponents_SF_full}.}
     \label{fig:degree_2CDCcore}
 \end{figure*}

\begin{figure*}[!htb]
    \centering
    \includegraphics[width=0.99\linewidth]{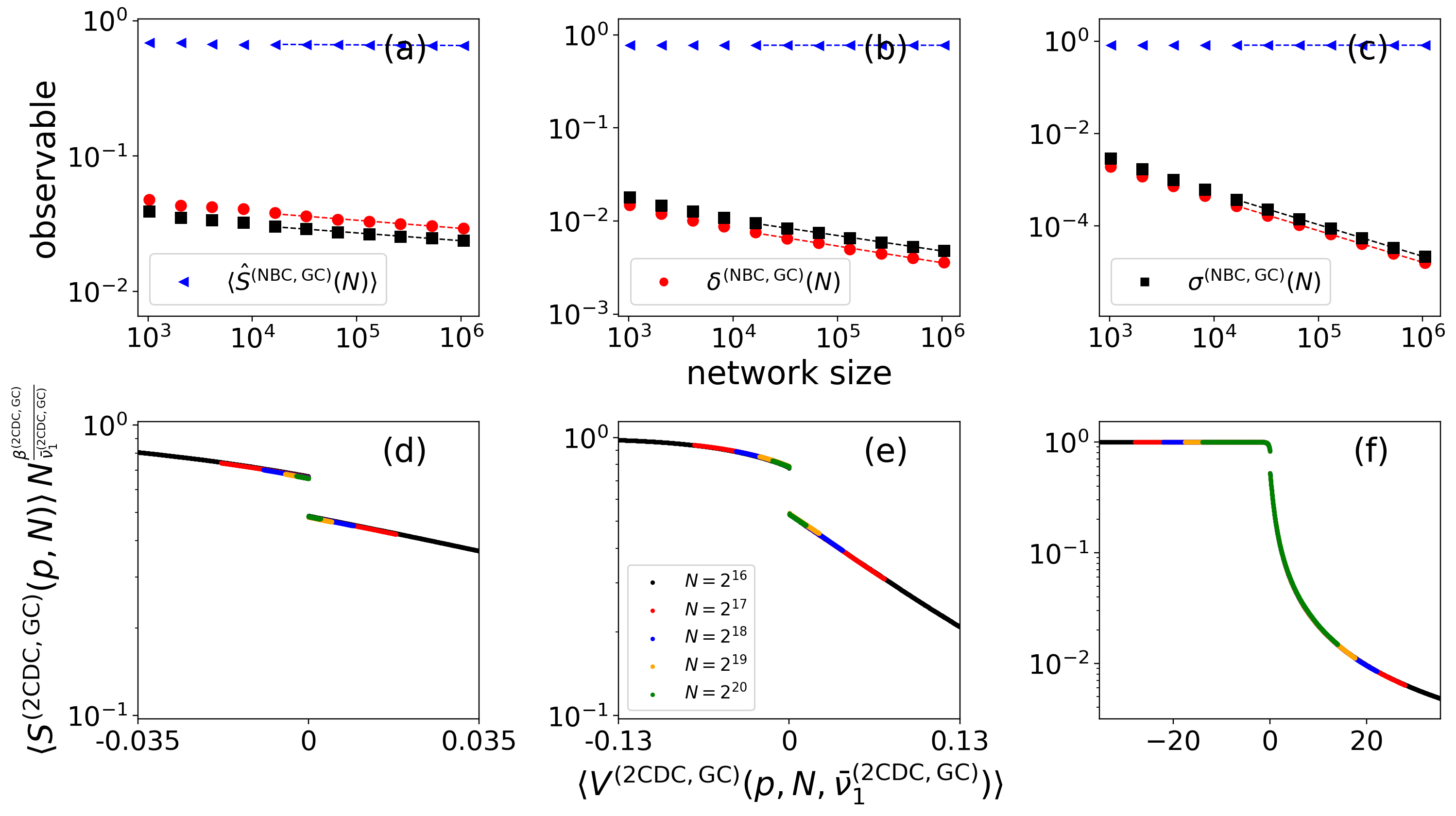}
    \caption{{\bf Critical properties of the GC transition for 2CDC-biased percolation on scale-free graphs.} 
    Same as in Figure~\ref{fig:2}, but for the 2CDC-biased percolation process applied to scale-free graphs with degree exponent $\gamma = 2.5$. 
     Results are averaged over $Q = 10000$ instances of the biased percolation model.
    Best estimates of the critical exponents are reported in Tab.~\ref{tab:exponents_SF_full}.}
\end{figure*}

\begin{figure*}[!htb]
    \centering
    \includegraphics[width=0.99\linewidth]{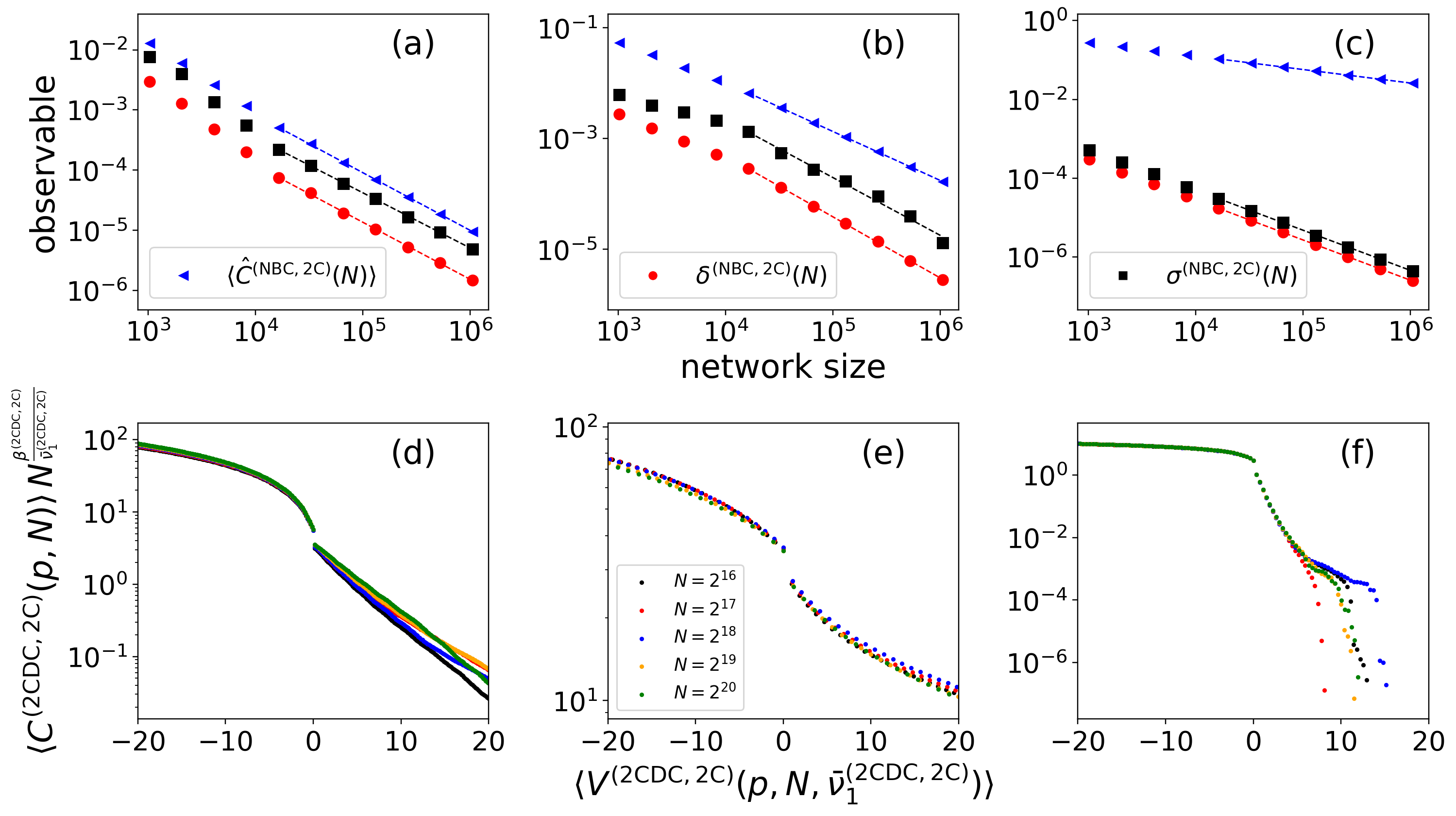}
    \caption{{\bf Critical properties of the 2C transition for 2CDC-biased percolation on scale-free graphs.} 
    Same as in Figure~\ref{fig:3}, but for the 2CDC-biased percolation process applied to scale-free graphs with degree exponent $\gamma = 2.5$. 
     Results are averaged over $Q = 10000$ instances of the biased percolation model.
    Best estimates of the critical exponents are reported in Tab.~\ref{tab:exponents_SF_full}.}
\end{figure*}

\begin{figure*}[!htb]
    \centering
    \includegraphics[width=0.99\linewidth]{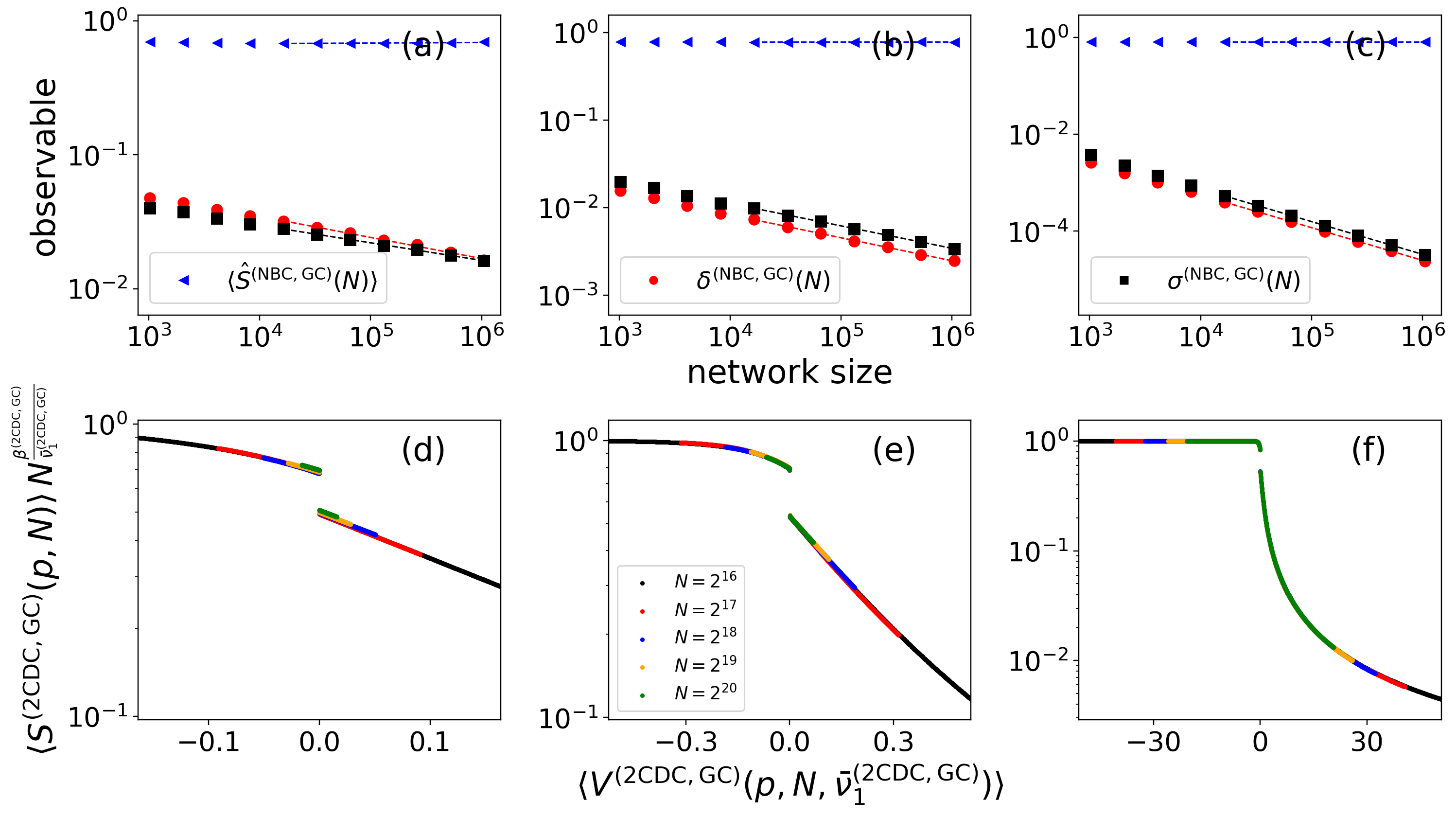}
    \caption{{\bf Critical properties of the GC transition for 2CDC-biased percolation on scale-free graphs.} 
    Same as in Figure~\ref{fig:2}, but for the 2CDC-biased percolation process applied to scale-free graphs with degree exponent $\gamma = 3.5$.  
     Results are averaged over $Q = 10000$ instances of the biased percolation model.
    Best estimates of the critical exponents are reported in Tab.~\ref{tab:exponents_SF_full}.}
\end{figure*}

\begin{figure*}[!htb]
    \centering
    \includegraphics[width=0.99\linewidth]{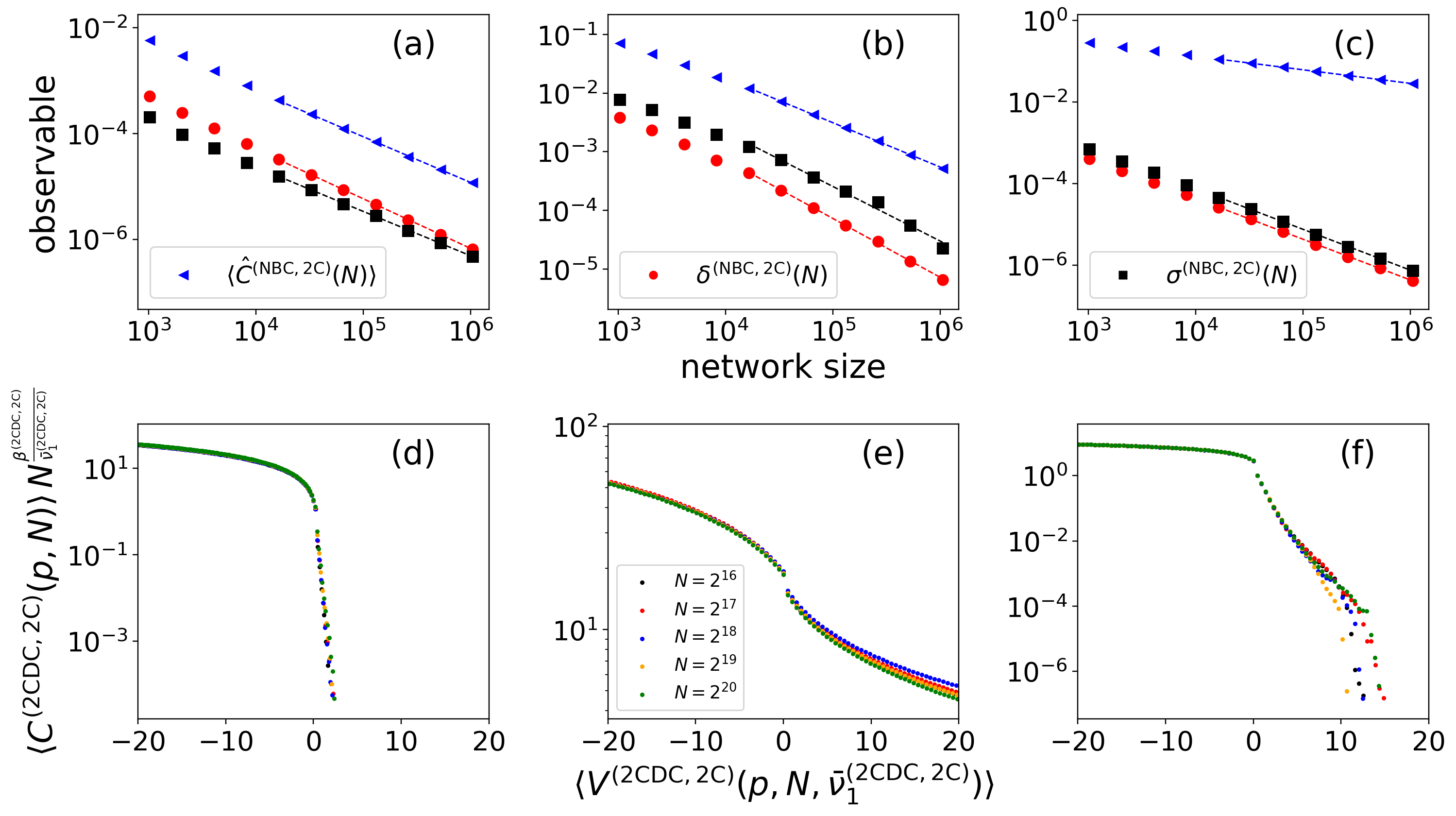}
    \caption{{\bf Critical properties of the 2C transition for 2CDC-biased percolation on scale-free graphs.} 
    Same as in Figure~\ref{fig:3}, but for the 2CDC-biased percolation process applied to scale-free graphs with degree exponent $\gamma = 3.5$.  
     Results are averaged over $Q = 10000$ instances of the biased percolation model.
    Best estimates of the critical exponents are reported in Tab.~\ref{tab:exponents_SF_full}.}
\end{figure*}

\end{document}